\documentclass[usenatbib]{mnras}

\usepackage[T1]{fontenc}

\usepackage{amsmath,txfonts}

\usepackage{amsmath}	
\usepackage{mathtools}  
\usepackage{graphicx}	
\usepackage[normalem]{ulem} 
\usepackage{multirow} 
\usepackage{xcolor} 
\usepackage{subcaption}
\usepackage{float}
\usepackage{siunitx}
\usepackage{booktabs}
\usepackage{algorithm} 
\usepackage{algpseudocode}
\usepackage{hyperref} 

\newcommand{\Lya}{Ly-$\alpha$}

\newcolumntype{L}[1]{>{\centering\arraybackslash$}p{#1}<{$}}
\renewcommand{\arraystretch}{1.2}

\DeclareRobustCommand{\VAN}[3]{#2}
\let\VANthebibliography\thebibliography
\def\thebibliography{\DeclareRobustCommand{\VAN}[3]{##3}\VANthebibliography}

\makeatletter
\def\footnoterule{\kern-3\p@
  \hrule \@width 2in \kern 2.6\p@} 
\makeatother
\title{Improved Halo Model Calibrations for Mixed Dark Matter Models of Ultralight Axions}

\author[T. Dome et al.]
{\parbox[t]{\textwidth}
{Tibor Dome$^{1,2}$\thanks{E-mail: td448@cam.ac.uk}, 
Simon May$^{3,4}$,
Alex Lagu\"{e}$^{5}$,
David J. E. Marsh$^{7}$,
Sarah Johnston$^{6}$,
Sownak Bose$^{6}$,
Alex Tocher$^{1,2}$, 
Anastasia Fialkov$^{1,2}$
}
\\ \\
$^{1}$Institute of Astronomy, University of Cambridge, Madingley Road, Cambridge, CB3 0HA, UK\\
$^{2}$Kavli Institute for Cosmology, Madingley Road, Cambridge, CB3 0HA, UK\\
$^{3}$Perimeter Institute for Theoretical Physics, 31 Caroline Street North, Waterloo, ON, N2L 2Y5, Canada\\
$^{4}$Department of Physics, North Carolina State University, Raleigh, NC, 27695-8202, USA\\
$^{5}$Department of Physics and Astronomy, University of Pennsylvania, 209 South 33rd Street, Philadelphia, PA, USA 19104\\
$^{6}$Institute for Computational Cosmology, Department of Physics, Durham University, South Road, Durham DH1 3LE, UK\\
$^{7}$King's College London, Strand, London, WC2R 2LS, United Kingdom
}

\date{Accepted XXX. Received YYY; in original form ZZZ}

\pubyear{2024}

\begin{document}
\label{firstpage}
\pagerange{\pageref{firstpage}--\pageref{lastpage}}
\maketitle

\begin{abstract}
We study the implications of relaxing the requirement for ultralight axions to account for all dark matter in the Universe by examining mixed dark matter (MDM) cosmologies with axion fractions $f \leq 0.3$ within the fuzzy dark matter (FDM) window $10^{-25}$\,eV $\lesssim m \lesssim 10^{-23}$\,eV. Our simulations, using a new MDM gravity solver implemented in \textsc{AxiREPO}, capture wave dynamics across various scales with high accuracy down to redshifts $z\approx 1$. We identify halos with \textsc{Rockstar} using the CDM component and find good agreement of inferred halo mass functions (HMFs) and concentration–mass relations with theoretical models across redshifts $z=1-10$. This justifies our halo finder approach a posteriori as well as the assumptions underlying the MDM halo model \textsc{AxionHMcode}. Using the inferred axion halo mass–cold halo mass relation $M_{\text{a}}(M_{\text{c}})$ and calibrating a generalised smoothing parameter $\alpha$ to our MDM simulations, we present a new version of \textsc{AxionHMcode}. The code exhibits excellent agreement with simulations on scales $k< 20 \, h \, \text{cMpc}^{-1}$ at redshifts $z=1-3.5$ for $f\leq 0.1$ around the fiducial axion mass $m = 10^{-24.5}\,\text{eV} = 3.16\times 10^{-25}\,\text{eV}$, with maximum deviations remaining below 10\,\%. For axion fractions $f\leq 0.3$, the model maintains accuracy with deviations under 20\,\% at redshifts $z\approx 1$ and scales $k< 10 \, h \, \text{cMpc}^{-1}$, though deviations can reach up to 30\,\% for higher redshifts when $f=0.3$. Reducing the run-time for a single evaluation of \textsc{AxionHMcode} to below $1$ minute, these results highlight the potential of \textsc{AxionHMcode} to provide a robust framework for parameter sampling across MDM cosmologies in Bayesian constraint and forecast analyses.
\end{abstract}




\section{Introduction}
\label{s_intro}
Ideal candidates for ultralight bosonic dark matter (DM) – also called fuzzy dark matter (FDM) – from particle physics are ultralight axions which can emerge from both field theory and string theory. In field theory, these candidates are typically axion-like particles (ALPs), which share their conceptual origins with quantum chromodynamics (QCD) axions \citep{Preskill_1983, Abbott_1983, Dine_1983} but can have much lighter masses with couplings to the Standard Model typically taken as free parameters \citep[although see][]{Dias_2014, Kim_2016}. In string theory, pseudoscalar fields with axion-like properties arise in compactifications as Kaluza–Klein (KK) zero modes of antisymmetric tensor fields, with their masses and couplings determined by the topology and geometry of the compact manifold \citep{Svrcek_2006,Arvanitaki_2010, Visinelli_2019, Gendler_2023}.

In both field theory and string theory, the lightest of these axions could have a mass lower than $10^{-19}$\,eV, which implies the existence of particles with de Broglie wavelengths on galactic or even cosmological scales. In the last decade, several extensions of the simplest one-field FDM model have been explored in the literature. Some notable studies include vector FDM where FDM is a higher-spin field \citep{Amin_2022}, multi-field FDM \citep{Gosenca_2023, Luu_2024}, and self-interacting FDM \citep{Mocz_2023} where the quartic self-coupling term is included. In the self-interacting model, soliton cores can undergo a phase transition from dilute to denser states, and small-scale structure formation may be enhanced.

String theory compactifications typically yield a plenitude of $N=\mathcal{O}(100)$ axion fields with logarithmically distributed masses \citep{Mehta_2021}. Importantly, the relic density of ultralight particles, $\Omega_{\text{FDM}}$, does not necessarily match the total DM density of the Universe, $\Omega_{\text{m}}-\Omega_{\text{b}}$. For example, \cite{Bachlechner_2017} show that in theories with $N=\mathcal{O}(100)$ axions and a lightest axion of mass $m\sim 10^{-22}$\,eV, axion misalignment can generate a DM abundance of $\Omega_{\text{FDM}} \sim 0.1$. Models with a small number $N=\mathcal{O}(1)$ of axions can also give rise to $\Omega_{\text{FDM}} \sim 0.1$ via axion misalignment \citep{Cicoli_2022_2} if the product $Sf_{\text{a}}$ of the instanton action $S$ (giving rise to the axion potential) and the axion decay constant $f_{\text{a}}$ satisfies $Sf_{\text{a}} \gtrsim 1$, slightly violating the Weak Gravity Conjecture \citep{Hebecker_2018, Alonso_2017}. This makes phenomenological multi-field FDM models a natural extension within the framework of string theory.

The phenomenology of the two-field FDM model can be complex, novel, and compelling, as shown by several studies. \cite{Huang_2023} conducted the first two-field cosmological simulations with cold dark matter (CDM) initial conditions and showed that minor admixtures of heavier axions to a lighter-mass major component can neither significantly disturb stable major-component solitons nor form minor-component solitons. \cite{Gosenca_2023} studied stellar heating in FDM halos with two or more fields, while \cite{Glennon_2023} simulated soliton collision of two-field FDM with inter-field and self-interactions. \cite{Jain_2024} investigated Bose-Einstein condensation (BEC) formation in the kinetic regime, and \cite{Luu_2024} showed that in the regime where the mass ratio $m_2/m_1$ is in the range $2-7$, a minimal model with uncoupled fields predicts a central soliton with a nested structure, distinguishable from the typical flat-core soliton found in one-field halos. Importantly, multi-field FDM circumvents the Catch 22 problem that challenges the use of the simplest non-interacting one-field FDM as a solution to the small scale problems of CDM \citep{Marsh_2014}.

In the context of mixed dark matter (MDM) models, we make the simplifying assumption that only one of the DM particle species is ultralight while the others have negligible de Broglie wavelengths. In this case, we group the combined relic density of the \mbox{(near-)}collisionless components into $\Omega_{\text{CDM}}$ since the species can be modelled as CDM by virtue of the correspondence between the Schrödinger–Poisson and Vlasov–Poisson equations \citep{Mocz_2018}. How do observations constrain MDM models? While the FDM constraints compiled in \cite{Dome_2022} assume a pure FDM cosmology, many constraints were recently reported assuming axions do not constitute all of the DM in the Universe. In the mass range $10^{-32}$\,eV $\lesssim m \lesssim 10^{-26}$\,eV, FDM can only comprise a few percent of the total DM \citep{Hlozek_2015, Hlozek_2018, Lague_2022, Rogers_2023}. For higher values of $m$, the bound becomes weaker since the FDM power suppression moves to smaller scales, while for smaller values of $m$, FDM behaves as dark energy and is strongly degenerate \mbox{with $\Omega_{\Lambda}$}.

At the higher-mass end, \Lya\ forest and ultra-faint dwarf (UFD) data indicates that DM cannot be fully described by pure FDM models in the mass range $10^{-21}$\,eV $\lesssim m \lesssim 3\times 10^{-19}$\,eV \citep{Armengaud_2017, Marsh_2019, Zimmermann_2021, Dalal_2022, Irsic_2023}, with a stronger bound for $10^{-23}$\,eV $\lesssim m \lesssim 10^{-21}$\,eV where FDM must not comprise more than $\mathcal{O}(10\,\%)$  of the total DM \citep{Kobayashi_2017}. The Dark Energy Survey year 1 was used by \cite{Dentler_2022} to search for shear-correlation suppressions caused by FDM (assuming pure FDM), ruling out the existence of pure FDM in the mass range $10^{-25}$\,eV $\lesssim m \lesssim 10^{-23}$\,eV \citep[also see][]{Preston_2024}. However, current constraints suggest that it is possible for FDM to exist in fairly large portions (albeit not the entirety of the DM) in the region $10^{-25}$\,eV $\lesssim m \lesssim 10^{-23}$\,eV, referred to as the \textit{FDM window}.

Recent studies have demonstrated the potential of 21 cm interferometers, particularly the Hydrogen Epoch of Reionisation Array (HERA), to detect FDM fractions of $f\gtrsim \mathcal{O}(1\,\%)$ within the FDM window and beyond \citep{Jones_2021, Flitter_2022}. This potential was further supported by \cite{Lazare_2024}, who in addition established an upper bound on FDM with a particle mass $m=10^{-23}$\,eV, restricting it to 16\,\% of the total DM. Their constraints are based on a range of observations, including UV luminosity function (LF) data from the Hubble Space Telescope, constraints on the neutral hydrogen fraction from high-redshift quasar spectroscopy, cosmic microwave background (CMB) optical depth measurements from Planck, and upper bounds on the 21 cm power spectrum from HERA. These constraints tighten for smaller masses, reaching down to 1\,\% for $m=10^{-26}$\,eV. However, their machine learning-based emulator does neither account for the `quantum pressure' term in the evolution of the axion field, nor the impact of FDM on star formation, and \textsc{21cmFAST} relies on the excursion set formalism \citep{Mesinger_2011}, which may not adequately capture small-scale physics. Using UVLF and Planck CMB data only, \cite{Winch_2024} report slightly weaker constraints, restricting FDM to less than $\approx 22\,\%$ of the DM at $m=10^{-23}$\,eV and to less than $\approx 5\,\%$ at $m=10^{-26}$\,eV.

The future Square Kilometre Array (SKA) is expected to provide strong constraints on FDM and MDM models, at few per cent level for masses $m\sim 10^{-22}$\,eV, and improving to below 1\,\% for lighter masses \citep{Bauer_2021, Dome_2024_2}. For pure FDM, \cite{Hotinli_2022} showed that by means of velocity acoustic oscillations in the large-scale 21\,cm power spectrum, HERA may be sensitive to axions with masses up to $m \approx 10^{-18}$\,eV \cite[see also][for FDM relative velocity effects]{Marsh_2015_2}.

\cite{Shevchuk_2023} argued that FDM with a particle mass of $m \lesssim 10^{-24}$\,eV is inconsistent with the observed Einstein radii of several strong lensing systems. However, their analysis is based on simplified models, such as expressing total DM density profiles as $\rho_{\text{DM}}(r) = f\rho_{\text{FDM}}(r)+(1-f)\rho_{\text{CDM}}(r)$, and assuming idealised solitons derived solely from FDM simulations, without accounting for contributions from cold dark matter or baryons.

We thus believe that the FDM window remains an appealing regime to explore in MDM simulations, for a number of reasons. First, \cite{Blum_2021} proposed that $\mathcal{O}(10\,\%)$ of DM in the form of FDM with particle mass in the window, $m \sim 10^{-25}$\,eV, could explain the tension between inferences of $H_0$, the current expansion rate of the Universe, which are based on the time delay in lensed quasar measurements, and those based on CMB observations. Note that both \cite{Shevchuk_2023} and \cite{Blum_2021} assume a soliton core in the (lens) density profile for FDM fractions of $\mathcal{O}(10\,\%)$, the validity of which we will address in more detail later (see Secs.~\ref{ss_cutoff} and \ref{ss_axionhalomass}). Second, the 5$\sigma$ tension between Planck CMB and \Lya\ forest data might prefer a fraction $\approx (1 - 5) \,\%$ of FDM with particle mass $m\sim 10^{-25}$\,eV, which is very close to the axion mass that we probe \citep{Rogers_2023}. Third, a fraction $f \sim10\,\%$ of FDM was suggested to explain the suppressed amplitude of the matter power spectrum at late times \citep[also known as the $\sigma_8$/S$_8$ tension,][]{Allali_2021, Ye_2023}.

In this work, our aim is to push both the numerical and \mbox{(semi-)}\allowbreak analytical frontiers of MDM modelling. Alongside implementing a new MDM gravity solver and conducting a suite of simulations within the FDM window for a particle mass of $m = 10^{-24.5}\,\text{eV} = 3.16\times 10^{-25}\,\text{eV}$, we refine the halo model framework \textsc{AxionHMcode} \citep{Vogt_2023}, calibrating some of its key parameters to the MDM simulations. We will achieve this mainly by modelling the axion halo mass–cold halo mass relation $M_{\text{a}}(M_{\text{c}})$ as a broken power law below the linear Jeans mass and calibrating a generalised transition smoothing parameter $\alpha$.

This paper is organised as follows. We start with the physics of MDM in Sec.~\ref{ss_mdm_phys} and describe the numerical methodology in Sec.~\ref{ss_pseudo}, which we implement in \textsc{AxiREPO} \citep{May_2021, May_2022}. Key findings regarding large-scale structure and halo statistics are presented in Sec.~\ref{s_mdm_stats}. We detail our improvements to MDM halo model calibrations in Sec.~\ref{s_calibrate} before we conclude in Sec.~\ref{s_mdm_outlook}.

\section{Methods}
\label{s_methods}

\subsection{Mixed Dark Matter Physics}
\label{ss_mdm_phys}
We define the FDM fraction (of the total matter density) as
\begin{equation}
f \equiv \frac{\Omega_{\text{FDM}}}{\Omega_{\text{FDM}} + \Omega_{\text{CDM}} + \Omega_{\text{b}}},
\label{e_fdef}
\end{equation}
where we adopt \cite{Planck_2015} cosmological parameters. In this paper, the simulation-based analyses (up to Sec.~\ref{s_calibrate}) rely on DM-only modelling, where baryons are effectively absorbed into the CDM component and assumed to behave similarly to CDM, such that $\Omega_{\text{CDM}} \leftarrow \Omega_{\text{CDM}} + \Omega_{\text{b}}$ and $\Omega_{\text{b}} \leftarrow 0$.
In other words, for the purposes of our simulations, we set $\Omega_{\text{b}} = 0$ and increase $\Omega_{\text{CDM}}$ correspondingly.

In DM-only models (as in our simulations) this definition of $f$ is standard \citep[e.\,g.][]{Lague_2023}, and $f_{\text{max}} = 1$ corresponds to pure FDM. In models that distinguish between the CDM component and baryons, our definition~\eqref{e_fdef} differs from the more commonly used $\tilde{f} = \Omega_{\text{FDM}} / (\Omega_{\text{FDM}} + \Omega_{\text{CDM}})$, which is also employed in the \cite{Vogt_2023} MDM halo model central to this paper. In such cases, the maximum FDM fraction is $f_{\text{max}} = 0.843$. Converting between the more common definition and Eq.~\eqref{e_fdef} simply involves downscaling $\tilde{f}$ by the factor $(\Omega_{\text{FDM}} + \Omega_{\text{CDM}}) / (\Omega_{\text{FDM}} + \Omega_{\text{CDM}} + \Omega_{\text{b}})$.

We consider axions and axion-like particles, characterised by a periodicity defined by the energy scale $f_{\text{a}}$ and represented as an angular variable $\phi$. The initial, random value of the misalignment angle $\theta_a = \phi/f_{\text{a}}$ (or the distribution thereof in our Hubble patch) determines the FDM abundance $\Omega_{\text{FDM}}$. In case of the QCD axion, the associated global $\mathrm{U}(1)$ symmetry is the Peccei–Quinn (PQ) symmetry whose angular pseudoscalar field $\phi$ couples to the strong sector and via non-perturbative QCD effects (instantons) dynamically relaxes the $\theta$ term in the QCD Lagrangian to zero, thus `solving' the strong-CP problem \citep{Peccei_1977}. Similar non-perturbative effects also exist in string-theoretical realisations of axions \citep[brane instantons,][]{Svrcek_2006, Svrcek_2006_2, Blumenhagen_2009} and in both cases lead to the spontaneous breaking of the exact shift symmetry $\phi \rightarrow \phi + \text{const.}$ at scale $\Lambda \ll f_{\text{a}}$, rendering axions and axion-like particles massive (hence the term pseudo-Nambu–Goldstone boson). The potential produced by the non-perturbative effects respects the periodicity of $\phi$ and is usually written as
\begin{equation}
V(\phi) = \Lambda^4\left[1-\cos\left(\frac{\phi}{f_{\text{a}}}\right)\right],
\end{equation}
where the axion mass can be expressed as $m = \Lambda^2/f_{\text{a}}$ in case of string theory models and using temperature-dependent terms in case of QCD axions \citep{Niemeyer_2020}.

In the limit of small field displacements away from the potential minimum, $\phi \ll f_{\text{a}}$, and ignoring the Hubble drag in a late-time Friedmann–Lemaître–Robertson–Walker (FLRW) background cosmology, the axion satisfies the Klein-Gordon equation (in natural units),
\begin{equation}
(-\partial_{\mu}\partial^{\mu}+m^2)\phi = 0.
\end{equation}
Here we are interested in the case where $\phi$ has both spatial and temporal fluctuations. On scales much larger than the Compton wavelength (in natural units)
\begin{equation}
\lambda_{\text{c}} \equiv \frac{2\pi}{m},
\label{e_compton}
\end{equation}
but much smaller than the particle horizon, one can employ a nonrelativistic approximation to the dispersion relation and a Newtonian approximation to the gravitational interaction embedded in
the covariant derivatives (since $\partial_{\mu}\partial^{\mu}=g^{\mu\nu}\partial_{\mu}\partial_{\nu}$) of the field equation.

It is then convenient to define a complex scalar field 
\begin{equation}
\psi = \sqrt{\rho_{\text{FDM}}}e^{i\gamma},
\label{e_classical_wf}
\end{equation}
the `classical' wavefunction, constructed from the amplitude and phase of the field $\phi$,
\begin{equation}
\phi = \sqrt{\rho_{\text{FDM}}}\cos (mt-\gamma).
\end{equation}
The wavefunction then obeys
\begin{subequations}\label{e_sp_proper}
\begin{align}
i\left(\partial_t+\frac{3}{2}\frac{\dot{a}}{a}\right)\psi &= \left(-\frac{1}{2m}\nabla^2+m\Psi\right)\psi, \\
\nabla^2\Psi &= 4\pi G (\rho_{\text{CDM}}+\rho_{\text{FDM}}-\bar{\rho}_{\text{tot}}),
\label{e_poisson_proper}
\end{align}
\end{subequations}
where $\Psi$ is the Newtonian gravitational potential, the FDM density can be expressed as $\rho_{\text{FDM}} = |\psi|^2$ and $\bar{\rho}_{\text{tot}}$ is the mean of the total DM density. This is simply the non-linear Schrödinger–Poisson (SP) system of equations for a self-gravitating many-body field in a potential well, embedded in an expanding Universe. Note that the right hand side of the first equation of system \eqref{e_sp_proper} vanishes for the unperturbed background and the energy density in the axion field $\rho_{\text{FDM}} =|\psi|^2\propto a^{-3}$ redshifts like matter. 

In terms of comoving coordinates $\mathbf{x}\equiv \mathbf{r}/a$, the SP equations become
\begin{subequations}\label{e_sp_comov}
\begin{align}
i\frac{\partial \psi_{\text{c}}}{\partial t} &= -a^{-2}\frac{1}{2m}\nabla^2_{\text{c}}\psi_{\text{c}} + a^{-1}m\Psi_{\text{c}}\psi_{\text{c}},\\
\nabla_{\text{c}}^2\Psi_{\text{c}} &= 4\pi G (\rho_{\text{CDM,c}}+\rho_{\text{FDM,c}}-\bar{\rho}_{\text{m,c}}),
\label{e_poisson_comoving}
\end{align}
\end{subequations}
where we have defined comoving quantities, relating to physical quantities as
\begin{equation}
\rho_{i, \text{c}}\equiv a^3\rho_i, \ \ \psi_{\text{c}}\equiv a^{3/2}\psi, \ \ \nabla_{\text{c}}\equiv a \nabla, \ \ \Psi_{\text{c}}\equiv a\Psi,
\end{equation}
for $i\in \lbrace \text{CDM}, \text{FDM}, \text{m}\rbrace$, where `m' denotes total matter. The background density of the Universe is $\bar{\rho}_{\text{m,c}} = \Omega_{\text{m}}\rho_{\text{crit}} = \bar{\rho}_0$. In a fully relativistic formulation, as derived from the linearized Einstein equations, the Poisson equations~\ref{e_poisson_proper} and \ref{e_poisson_comoving} contain an $a^2$ factor on the right-hand side \citep[see, e.g.,][]{Mukhanov_2005, Marsh_2016, Baumann_2022}. However, here we adopt the Newtonian approximation, which applies on scales much smaller than the Hubble horizon and treats perturbations as quasi-static. As such, the factor of $a^2$ is omitted, consistent with previous works studying scalar field dark matter in this regime \citep[e.g.,][]{Schive_2014, Hui_2021}.

\subsection{Characteristic Scales in MDM}
\label{ss_4scales}
A fundamental length scale in the context of MDM is the comoving linear Jeans scale \citep{Hu_2000, Marsh_2014},
\begin{align}
k_{\text{J}} &= \left(16\pi G a^4 \bar{\rho}_{\text{m}}(a) \right)^{1/4}\left(\frac{m}{\hbar}\right)^{1/2}\nonumber \\
&\simeq 66.5 \, (1+z)^{-1/4}\left(\frac{\Omega_{\text{m}}h^2}{0.12}\right)^{1/4} \left(\frac{m}{10^{-22} \, \text{eV}}\right)^{1/2} \text{cMpc}^{-1},
\label{e_jeans_fdm}
\end{align}
where $\bar{\rho}_{\text{m}}(a)$ is the scale factor dependent background density in physical units. The scale $k_{\text{J}}$ emerges from the non-vanishing effective sound speed of scalar fields, which introduces a Jeans-like dispersion relation for perturbation modes. Notably, the linear Jeans scale exhibits minimal dependence on redshift and is unaffected by the axion fraction $f$. The corresponding linear Jeans mass can be written as
\begin{equation}
M_{\text{J}} = \frac{4}{3}\pi\left(\frac{\pi}{k_{\text{J}}}\right)^3\bar{\rho}_{0} \approx (1-2) \times 10^{11} \, M_{\odot}/h \quad \text{for} \ \ z=1-4,
\label{e_Jeansmass}
\end{equation}
where we have assumed spherical basis functions, and as usual we have taken $m=3.16\times 10^{-25}$\,eV.\par

In pure FDM cosmologies ($f=f_{\text{max}}$), no halos can form with mass $M < M_{\text{J}}$ due to the pressure support from the scalar field. Some residual structures (not halos) may still form via fragmentation. However, this estimate is based on a purely linear effect and indeed the growth rate is enhanced at second order \citep{Li_2019}, hence heuristic reasoning based on experience from the baryonic Jeans length can only be applied with caution. \cite{Schive_2014_2} derived a slightly more accurate minimum halo mass from the properties of FDM solitons that can form, but their estimate is in close agreement with the linear Jeans mass $M_{\text{J}}$.

\begin{figure*}
\includegraphics[width=\textwidth]{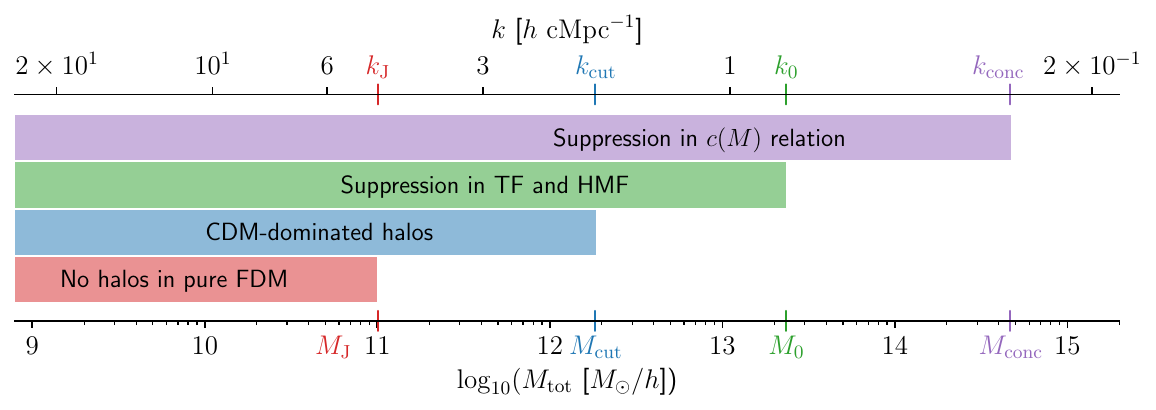}
\caption{Characteristic scales in the MDM framework. We illustrate the linear Jeans mass $M_{\text{J}}$ (see Eq.~\ref{e_jeans_fdm}), the cut-off mass $M_{\text{cut}}$ (see Eq.~\ref{e_rhalo_jeans_length} and associated text), and the characteristic mass $M_0$, where deviations from CDM become apparent in the matter transfer function (TF) and HMF (see Eq.~\ref{e_M0}). Additionally, we identify $M_{\text{conc}}$, the mass scale at which the concentration–mass relation $c(M)$ shows notable changes. Results are presented for a fiducial MDM cosmology with axion mass $m=3.16\times 10^{-25}$\,eV and axion fraction $f=0.1$ at redshift $z=1$.}
\label{f_4scales}
\end{figure*}

\begin{table}
	\centering
    \begin{tabular}{c c c c c }
     \toprule
     Variable & $k_{\text{cut}}$ & $k_{\text{J}}$ & $k_{0}$ & $k_{\text{conc}}$\\
     \midrule
     $m$ & $\phantom{-}0.66$ & $\phantom{-}1/2$ & $4/9$ & $4/9$ \\
     $1+z$ & $-0.24$ & $-1/4$ & $0$ & $0$ \\
     $f$ & $\phantom{-}0\phantom{.00}$ & $0$ & $0$ & $0$ \\
     \bottomrule
    \end{tabular}
    \caption{Scaling of several characteristic MDM scales with axion mass $m$, redshift $z$, and axion fraction $f$. The best-fit results for the scaling of the cut-off scale $k_{\text{cut}}$ with axion mass, redshift, and axion fraction were obtained by evaluating Eq.~\eqref{e_rhalo_jeans_length} and comparing it to the virial radius in the redshift range $z=1-4$, axion mass range $m=10^{-26}-10^{-23}$\,eV, and axion fraction range $f=0.01-0.3$, respectively. These length scales show no dependence on the axion fraction $f$ and are useful scales to consider for both FDM and MDM cosmologies.}
    \label{t_scaling}
\end{table}

The Jeans mass can be generalised to non-linear structures, by adding the dependence on the halo profile $\rho_{\text{NFW}}$ of the CDM halo. \cite{Hu_2000} and \cite{Marsh_2014} showed that in a pure axion cosmology, no virialised halo is expected to form if this so-called \textit{halo Jeans scale} $r_{\text{hJ}}$ is larger than the (comoving) virial radius $R_{\text{vir}}$. The corresponding mass scale is called the cut-off mass $M_{\text{cut}}$. In comoving units, the halo Jeans wavenumber is given by (cf. Eq.~\ref{e_jeans_fdm})
\begin{equation}
k_{\text{hJ}} = 66.5 \, (1+z)^{-1/4}\left(\frac{\Omega_{\text{m}}h^2}{0.12}\right)^{1/4}\left(\frac{m}{10^{-22} \, \text{eV}}\right)^{1/2}\left(\frac{\rho_{\text{NFW}}(r_{\text{hJ}})}{\bar{\rho}_{\text{m}}}\right)^{1/4} \!\! \text{cMpc}^{-1}.
\label{e_khalo_jeans_length}
\end{equation}
Here, $k_{\text{hJ}}$ is calculated by converting $r_{\text{hJ}} =\pi/k_{\text{hJ}}$ as in \cite{Marsh_2014} instead of $r_{\text{hJ}} =2\pi/k_{\text{hJ}}$ as in \cite{Hu_2000}. Since we assume that $r_{\text{hJ}}\leq R_{\text{vir}}$, the Navarro–Frenk–White (NFW) density at $r_{\text{hJ}}$ can be written as \citep{Navarro_1997}
\begin{equation}
\rho_{\text{NFW}}(r_{\text{hJ}}) = \frac{\bar{\rho}_{\text{m}}\Delta_{\text{vir}}c^2}{3f(c)}\frac{R_{\text{vir}}}{r_{\text{hJ}}(1+r_{\text{hJ}}c/R_{\text{vir}})^2},
\label{e_rhoNFW}
\end{equation}
with $f(x) = -x/(1+x)+\ln(1+x)$ and $c = R_{\text{vir}}/r_{-2}$ being the halo concentration. The virial mass definition we adopt is
\begin{equation}
M_{\text{vir}} = M_{\text{tot}} = \frac{4\pi}{3}\bar{\rho}_0 R_{\text{vir}}^3\Delta_{\text{vir}}(z),
\end{equation}
where $\Delta_{\text{vir}}$ is the virial overdensity \citep{Bryan_1998}. Substituting Eq.~\eqref{e_rhoNFW} back into Eq.~\eqref{e_khalo_jeans_length}, we obtain
\begin{align}
r_{\text{hJ}}^3 = \left(\frac{\pi}{66.5}\right)^4 (1+z) \left(\frac{m}{10^{-22} \, \text{eV}}\right)^{-2}\left(\frac{\Omega_{\text{m}}h^2}{0.12}\right)^{-1}\frac{3f(c)}{\Delta_{\text{vir}}c^2}\frac{(1+r_{\text{hJ}}c/R_{\text{vir}})^2}{R_{\text{vir}}},
\label{e_rhalo_jeans_length}
\end{align}
where $r_{\text{hJ}}$ and $R_{\text{vir}}$ are measured in ckpc and we have assumed spherical basis functions as we did for Eq.~\eqref{e_Jeansmass}. Eq.~\eqref{e_rhalo_jeans_length} can then be solved for $r_{\text{hJ}}$. The dependence on axion fraction $f$ and redshift $z$ only enters via the concentration $c$ which we take to be the mean cosmic concentration at mass $M_{\text{vir}}$ as per Eqs.~\eqref{e_l16_model}. In fact, after comparing to the virial radius, we find that the cut-off mass is effectively independent of $f$, and attains values $M_{\text{cut}}\approx 1.7 \times 10^{12}-3.1\times 10^{13} \, M_{\odot}/h$ for our MDM cosmologies across redshifts $z=1-4$. 

How do we interpret the cut-off mass $M_{\text{cut}}$? In pure FDM, while the Jeans mass $M_{\text{J}}$ provides a fundamental lower bound on the mass of any halo that can form based on the balance between gravity and `quantum pressure' in the linear regime, the cut-off mass is a halo-specific lower bound that describes the minimum mass of virialised halos that can form. In MDM cosmologies, virialised halos can form below $M_{\text{cut}}$, but they will be primarily CDM-dominated.

Another important scale is the characteristic mass at which the non-linear matter power spectra and HMFs begin to deviate from the predictions of the CDM model. This characteristic mass, denoted as $M_0$, is given by \citep{Schive_2016}
\begin{equation}
M_0 = 1.6\times 10^{10}(m/10^{-22} \, \text{eV})^{-4/3} \, M_{\odot}
\label{e_M0}
\end{equation}
The scaling of $M_0$ is expected to be almost independent of redshift, as it is primarily set during the radiation-dominated epoch \citep{Hu_2000}.

Finally, we highlight a fourth mass scale, $M_{\text{conc}} \approx 20-50 \ M_0$, at which the mean halo concentration–mass relation $c(M)$ is significantly affected by axion physics. This was demonstrated by \cite{Bose_2015,Ludlow_2016,Dentler_2022} and can be attributed to the fact that halo concentration is determined when only a small fraction ($\approx 0.01$) of the mass has accumulated \citep{Navarro_1997, Bullock_2001}. These four mass scales are illustrated in Fig.~\ref{f_4scales}, using a fiducial MDM cosmology with axion mass $m=3.16\times 10^{-25}$\,eV and axion fraction $f=0.1$ at redshift $z=1$.

For reference, we show the scalings of these four characteristic length scales with axion mass $m$, redshift $z$ and axion fraction $f$ in Table~\ref{t_scaling}. We translate between mass and inverse length scales via $M=4\pi\bar{\rho}_0 (\pi/k)^3/3$. Among the four scales, the cut-off scale $k_{\text{cut}}$ shows the strongest dependence on axion mass, scaling as $k_{\text{cut}}\propto m^{0.66}$. Even though these scales were first introduced for pure FDM cosmologies and thus have no dependence on the axion fraction (while $M_{\text{cut}}$ is effectively independent of $f$), they remain valuable to consider in the context of MDM cosmologies as well. Specifically, we will show in Sec.~\ref{ss_hmfs} that despite its weak dependence on the axion fraction, $M_{\text{0}}$ serves as an effective scale for characterising the suppression of the halo mass function (HMF) at the low-mass end. In Sec.~\ref{ss_cM_relation}, we will show that the weak dependence of $M_{\text{conc}}$ on the axion fraction is corroborated by inferred concentration–mass relations. In addition, we will establish in Sec.~\ref{ss_axionhalomass} that $M_{\text{cut}}$ provides a useful approximate scale below which halos are predominantly CDM-dominated.

\subsection{Initial Conditions}
\label{ss_ics}
The goal of an initial conditions (ICs) generator for a cosmological simulation is to faithfully reproduce the statistical properties of the density field in the early universe with a finite number of point particles (for the CDM component) or grid cells (for the FDM component). Given a pre-IC particle distribution representing a completely homogeneous universe \cite[we use grid pre-ICs, see][]{Dome_2022}, the next task is to perturb the particles to produce a density field that reproduces the cosmologically relevant expected statistical properties. Let us consider a (Gaussian) over-density field $\delta(\mathbf{r})$ that is completely described by its power spectrum $P(k)=\langle \delta_{\mathbf{k}}^{} \delta_{\mathbf{k}}^{\ast}\rangle$. It is customary to express the amplitude of density fluctuations in terms of the transfer function $T(k,a)$,
\begin{equation}
P(k,a) \approx B k^{n_s} T(k,a)^2,
\end{equation}
where $n_{\text{s}}$ is the spectral index, and $B$ can be expressed in terms of the normalisation $A_{\text{s}}$ and the pivot scale $k_{\text{piv}}$ \citep{Planck_2015}.

Setting up ICs for cosmological simulations at a certain scale factor $a$ thus involves generating a white noise sample of random values $\mu(r)$ (typically sampled from a Gaussian distribution with zero mean and unit variance) and requiring that their amplitudes follow a specific power spectrum $P(k,a)$. This is achieved by multiplying the Fourier transformed white noise field $\mu_{\mathbf{k}}$ with the square root of the power spectrum, i.\,e. for all $\mathbf{k}$ representable on a grid of given resolution set
\begin{equation}
\delta_{\mathbf{k}}=\sqrt{P(k,a)} \, \mu_{\mathbf{k}}
= \sqrt{B} \,  k^{n_s/2} T(k,a) \mu_{\mathbf{k}}.
\label{e_deltak_white_noise}
\end{equation}
The real-space over-density field $\delta(\mathbf{r})$ is then obtained by inverse Fourier transformation, and this procedure is typically called ``$k$-space sampling''.

Note that a product in Fourier space simply corresponds to a convolution in real space, i.\,e. Eq.~\eqref{e_deltak_white_noise} is equivalent to
\begin{equation}
\delta(\mathbf{r})=\mathcal{T}(r,a)\star\mu(\mathbf{r}),
\label{e_deltar_white_noise}
\end{equation}
where $\mathcal{T}(r,a)$ is the real-space counterpart of $\mathcal{T}(k,a) = \sqrt{B} \, k^{n_s/2} T(k,a)$, and ``$\star$'' denotes a convolution. It is hence mathematically equivalent whether Eq.~\eqref{e_deltak_white_noise} is evaluated in Fourier space, followed by an inverse transform ($k$-space sampling), or whether Eq.~\eqref{e_deltar_white_noise} is evaluated using an inverse transform of $\mathcal{T}(k,a)$ followed by the convolution (real-space sampling). Most cosmological IC codes follow the first approach \citep[see e.\,g.][]{Bertschinger_2001}, while e.\,g. \cite{Pen_1997} and \cite{Sirko_2005} use the second or variations thereof.

However, the discrete realisations of the density fields derived with the two approaches will have \textit{significant} differences. This has been demonstrated conclusively by \cite{Sirko_2005}, who showed amongst others that employing Eq.~\eqref{e_deltak_white_noise} imposes periodicity of the real-space transfer function on box scales and leads to an underestimation of the two-point correlation function on large (sub-box) scales. This discrepancy arises because, with a finite number of particles in a finite box, achieving high accuracy for estimates of both the correlation function and the power spectrum is challenging due to finite box effects and the limitations in resolving all relevant scales. Even with an infinite number of particles, the finite volume of the box introduces periodic boundary conditions and aliasing effects that can distort the two-point correlation function, particularly at large scales. Importantly, in this work we use the public code \textsc{Music} to generate ICs \citep{Hahn_2011}, which is likewise based on a convolution of Gaussian white noise with a real-space transfer function kernel. The linear MDM power spectrum which serves as an input to \textsc{Music} is calculated using \textsc{AxionCamb} \citep{Hlozek_2015}.

Having generated the Eulerian-grid \textit{seed density field} $\delta(\mathbf{r})$ using real-space sampling while ensuring its consistency with both the input correlation function and the input power spectrum, we need to calculate the initial positions and velocities of the IC macroparticles. This is typically achieved using first (1LPT) or second order Lagrangian perturbation theory (2LPT), where $\delta(\mathbf{r})$ is used as the source field for Lagrangian perturbation theory. The displacement field in 1LPT only contains contributions from the gravitational potential, which is often called the Zel'dovich approximation \citep{Zeldovich_1970}. For all the simulations analysed in this work, we employ a more accurate representation by incorporating second-order effects using \textsc{Music}.

How do we take account of the multi-fluid nature of MDM beyond calculating separate density transfer functions for each component and coupling them gravitationally in the 2LPT? The growth of density perturbations in a two-component fluid can only be correctly reproduced if besides the different initial amplitudes of density perturbations also the difference in initial velocities between the two components are respected \citep{Yoshida_2003}. ICs for the two-component fluid thus ought to reflect these important differences between the two components.\footnote{This also applies to hydrodynamic simulations which simulate the evolution of baryons in addition to DM, not just MDM.} The correct growth of fluctuations in both components consistent with the predictions from linear perturbation theories can only be achieved with a proper modelling of (relative) velocities \citep{Somogyi_2010, Lague_2021}.

To obtain a velocity transfer function, recall that the curl of a peculiar velocity field, $\nabla \times \mathbf{v} \propto a^{-1}$, drops off with the expansion of the Universe and can be neglected at late times since there is no source for vorticity \citep[although see][]{Mocz_2017, Hui_2021_2}. Hence we can write $\mathbf{v}$ as the gradient of a velocity potential, $\mathbf{v}=\nabla \mathcal{V}$, and so
\begin{equation}
\mathbf{v}_{\mathbf{k}}=i\mathbf{k}\mathcal{V}_{\mathbf{k}}.
\end{equation}
Linear Newtonian perturbation theory applied to a pure DM cosmology (either CDM or FDM) thus implies that
\begin{equation}
\mathbf{v}_{\mathbf{k}}=\frac{ia\mathbf{k}}{k^2}\frac{\mathrm{d}\delta_{\mathbf{k}}}{\mathrm{d}t}.
\label{e_vk_def}
\end{equation}
For the growing mode of $\delta$, we obtain
\begin{equation}
\mathbf{v}_{\mathbf{k}}=\frac{ia\mathbf{k}}{k^2}Ha\delta_{\mathbf{k}}F(\Omega_{\text{m}}),
\label{e_vk_D}
\end{equation}
where
\begin{equation}
F(\Omega_{\text{m}})\equiv \frac{\mathrm{d}\ln(D_{+}(k,a))}{\mathrm{d}\ln(a)}.
\end{equation}
Note that Eq.~\eqref{e_vk_D} remains valid even in MDM cosmologies, where we have one such equation for each DM component. In contrast to pure CDM, the FDM growth rate depends on scale $k$, which arises due to the non-vanishing sound speed of the axion fluid \citep{Hwang_2009, Marsh_2016}.

\begin{figure}
\hspace*{-0.4cm}
\includegraphics[width=0.5\textwidth]{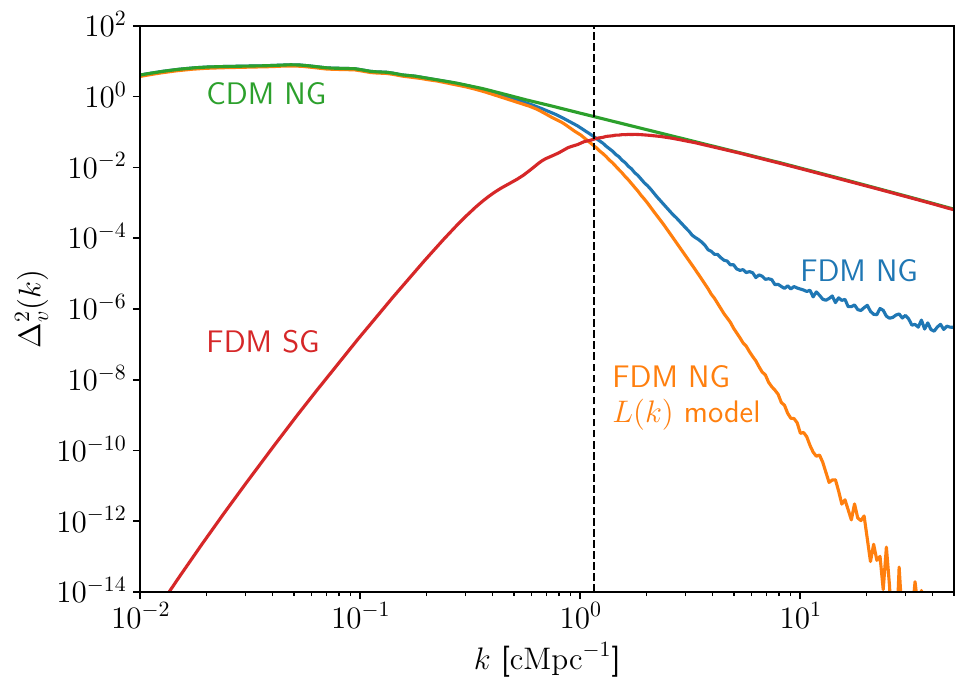}
\caption[Velocity transfer functions (squared) in an MDM cosmology consisting of a majority CDM component with an $f=0.1$ axion admixture of boson mass $m=10^{-24.5}$\,eV at $z=127$.]{Velocity transfer functions (squared) in an MDM cosmology consisting of a majority CDM component with an $f=0.1$ axion admixture of boson mass $m=10^{-24.5}$\,eV at $z=127$. The unapproximated FDM (blue solid) and CDM (green solid) velocity transfer functions in Newtonian gauge (NG) are calculated using a \textit{modified version} of \textsc{AxionCamb}. The vertical dashed line marks the FDM linear Jeans scale $k_{\text{J}}$ (Eq.~\ref{e_jeans_fdm}). The approximated Newtonian gauge FDM velocity, derived from MDM matter transfer functions (see Eq.~\ref{e_Lk_model}), is represented by the orange solid line. This $L(k)$ model provides a poor approximation to the blue solid line on small scales $k>5 \, \text{cMpc}^{-1}$. For completeness, we also show the (unapproximated) FDM velocity transfer function in synchronous gauge (SG), calculated similarly using a modified version of \textsc{AxionCamb}. Note that CDM SG is zero by definition, and so does not appear.}
\label{f_Lk_approx}
\end{figure}

It is nonetheless possible to approximately separate out the dependence on scale $k$ following \cite{Lague_2021},
\begin{equation}
D^{\text{FDM}}_{+}(k,a)\approx L(k)D^{\text{CDM}}_{+}(a).
\label{e_Lk_model}
\end{equation}
Note that we can express $L(k)$ as
\begin{equation}
L(k)\approx \frac{D^{\text{FDM}}_{+}(k,a)}{D^{\text{CDM}}_{+}(a)}=\sqrt{\frac{\Delta^{\text{FDM}}(k,a)}{\Delta^{\text{CDM}}(k,a)}},
\label{e_Lk_express}
\end{equation} 
where $\Delta(k,a) \equiv T(k,a)^2$ is the squared transfer function of the respective DM component. In CDM, recall that $T(k,a)=T(k)D_{+}(a)$.

We can now approximate the velocity field of the axion component in MDM cosmologies using the CDM velocity field as
\begin{equation}
\mathbf{v}^{\text{FDM}}_{\mathbf{k}} \approx L(k)\mathbf{v}^{\text{CDM}}_{\mathbf{k}}.
\end{equation}
To approximate the squared velocity transfer function $\Delta_{v}^{\text{FDM}}(k,a)$ for the axion component in an MDM cosmology, it should thus be sufficient to calculate $\Delta_{v}^{\text{CDM}}(k,a)$ in the corresponding pure CDM cosmology and construct the squared ratio of FDM-to-CDM matter transfer functions $\Delta^{\text{FDM}}(k,a)/\Delta^{\text{CDM}}(k,a)$. Assuming separability of $D^{\text{FDM}}_{+}(k,a)$ based on Eq.~\eqref{e_Lk_express}, we have
\begin{equation}
\Delta_{v}^{\text{FDM}}(k,a) \approx \frac{\Delta^{\text{FDM}}(k,a)}{\Delta^{\text{CDM}}(k,a)}\Delta_{v}^{\text{CDM}}(k,a).
\label{e_vk_approx}
\end{equation}
Now, both $\Delta_{v}^{\text{CDM}}(k,a)$ and the squared transfer function ratio can be generated easily using the public version of \textsc{AxionCamb}  \citep{Hlozek_2015}. We call Eq.~\eqref{e_vk_approx} the $L(k)$ model since it relies on the assumption of separability. To assess the fidelity of the $L(k)$ model, we have modified \textsc{AxionCamb} such that we can retrieve the unapproximated FDM velocity transfer function $\Delta_{v}^{\text{FDM}}(k,a)$ in Newtonian gauge directly from the Boltzmann code. In Fig.~\ref{f_Lk_approx}, we show velocity transfer functions in an MDM cosmology with a 10\,\% admixture of a $m=10^{-24.5}$\,eV axion field. We see that the $L(k)$ model provides a poor approximation for $\Delta_{v}^{\text{FDM}}(k,a)$ on small scales $k>5 \, \text{cMpc}^{-1}$ in Newtonian gauge,\footnote{On the other hand, in pure FDM cosmologies, the $L(k)$ model in fact provides a good approximation for $\Delta_{v}^{\text{FDM}}(k,a)$, and only on the smallest scales do we find deviations from the unapproximated result (not shown).} justifying our more reliable, unapproximated approach to velocity transfer functions.

Equipped with the velocity and matter transfer functions of each component, we use the public code \textsc{Music} to generate ICs \citep{Hahn_2011}. \textsc{Music} takes the transfer functions at a particular redshift (in our case, $z=127$) as input, and generates initial positions and velocities for macroparticles of each component.

\subsection{Madelung Formulation}
How do we initialise the FDM wavefunction on a grid from macroparticle positions and velocities? We use the Madelung change of variables \citep{Madelung_1927}, which can aid with physical intuition. We start with the decomposition \eqref{e_classical_wf} of the wavefunction into its amplitude and phase, $\psi_{\text{c}} = \sqrt{\rho_{\text{FDM,c}}}e^{i\gamma}$, and define the Madelung velocity as the gradient of the phase,
\begin{equation}
\mathbf{v}_{\text{M}}\equiv \frac{\nabla \gamma}{m}.
\label{e_mad_phase_vel}
\end{equation}
On linear scales before shell crossing, the fluid velocity is a gradient flow, and it resembles that of a superfluid. The Schrödinger equation can then be written as
\begin{subequations}
\begin{align}
\label{e_mass_conservation}
\frac{\partial \rho_{\text{FDM,c}}}{\partial t} + &\nabla_{\text{c}}\cdot (\rho_{\text{FDM,c}}\mathbf{v}_{\text{M}})=0,\\
\label{e_euler_equ_mad}
\frac{\partial\mathbf{v}_{\text{M}}}{\partial t}+a^{-2}\mathbf{v}_{\text{M}}\cdot \nabla_{\text{c}}\mathbf{v}_{\text{M}}=&-a^{-1}\nabla_{\text{c}}\Psi_{\text{c}}+a^{-2}\frac{1}{2m^2}\nabla_{\text{c}}\left(\frac{\nabla^2 \sqrt{\rho_{\text{FDM,c}}}}{\sqrt{\rho_{\text{FDM,c}}}}\right).
\end{align}
\end{subequations}
The Schrödinger equation possesses a $\mathrm{U}(1)$ symmetry, which amounts to the rotation of $\psi$ by a phase. With the identification of the fluid velocity, what is normally understood as probability conservation (i.\,e. conservation of the associated Noether current) in quantum mechanics is recast as mass conservation in Eq.~\eqref{e_mass_conservation}. The last term in the Euler equation \eqref{e_euler_equ_mad} is often referred to as the `quantum pressure' term. It is a misnomer since we have a classical system. In addition, the term arises from a stress tensor rather than mere pressure:
\begin{equation}
\Sigma_{ij}=\frac{1}{4m^2}\left(\rho^{-1}\partial_i\rho\partial_j\rho-\partial_i\partial_j\rho\right)=-\frac{\rho}{4m^2}\partial_i\partial_j\ln(\rho),
\label{e_stresstensor}
\end{equation}
i.\,e. $\partial_i(\nabla^2 \! \sqrt{\rho} \ /\sqrt{\rho})/(2m^2)=-\rho^{-1}\partial_j\Sigma_{ij}$. We have dropped the subscripts on the FDM density field in Eq.~\eqref{e_stresstensor}, $\rho = \rho_{\text{FDM,c}}$. The stress tensor $\Sigma_{ij}$ represents how the fluid description accounts for the underlying wave dynamics. It shows how the particle limit is obtained: for large $m$, the Euler equation reduces to that for a pressureless fluid, as is appropriate for particle DM. The insight that the wave formulation in the large $m$ limit can be used to model particle CDM was exploited by \cite{Widrow_1993}. The wave description effectively reshuffles information in a phase-space Boltzmann distribution into a position-space wavefunction and offers a number of insights that might otherwise be obscured \citep{Uhlemann_2019, Garny_2020}. This correspondence can be formalised \citep{Mocz_2018}. It is worth noting that \textsc{AxionCamb} also employs the Madelung description.

To initialise the FDM wavefunction, we calculate the FDM phase $\gamma$ by constructing the Madelung velocity field $\mathbf{v}_{\text{M}}$ and solving Eq.~\eqref{e_mad_phase_vel} in Fourier space. Equipped with CDM macroparticle positions and velocities as well as the FDM wavefunction, we can now evolve the joint MDM field.

\subsection{Pseudo-Spectral Method}
\label{ss_pseudo}
We use a spectral method to simulate MDM structure formation implemented in the \textsc{AxiREPO} code \citep{May_2021, May_2022}. The system of equations \eqref{e_sp_comov} is solved using a second-order symmetrised split-step pseudo-spectral Fourier method, colloquially called a `kick-drift-kick' leapfrog-like scheme. For a small time step $\Delta t$, the time evolution can be simplified using the following approximation \citep{Edwards_2018,May_2021}:
\begin{align}
&\psi_{\text{c}}(t+\Delta t, \mathbf{x})\nonumber\\
&= \mathcal{T}\exp\left[-i\int_t^{t+\Delta t}\left(-\frac{\hbar}{2m}\frac{1}{a(t')^2}\nabla^2_{\text{c}}+\frac{m}{\hbar}\frac{1}{a(t')}\Psi_{\text{c}}(t',\mathbf{x})\right)\mathrm{d}t'\right]\psi_{\text{c}}(t,\mathbf{x})\nonumber\\
&\approx \exp\left[i\frac{\Delta t}{2}\left(-\frac{\hbar}{m}\frac{1}{a(t)^2}\nabla^2_{\text{c}}-\frac{m}{\hbar}\frac{1}{a(t)}\Psi_{\text{c}}(t+\Delta t,\mathbf{x}) - \frac{m}{\hbar}\frac{1}{a(t)}\Psi_{\text{c}}(t,\mathbf{x}) \right)\right] \times\nonumber\\
&\qquad \psi_{\text{c}}(t, \mathbf{x})\nonumber\\
&\approx\exp\left[-i\frac{m}{\hbar}\frac{1}{a(t)}\frac{\Delta t}{2}\Psi_{\text{c}}(t+\Delta t,\mathbf{x})\right]\exp\left[i\frac{\hbar}{m}\frac{1}{a(t)^2}\frac{\Delta t}{2}\nabla^2_{\text{c}}\right]\times\nonumber\\
&\qquad\exp\left[-i\frac{m}{\hbar}\frac{1}{a(t)}\frac{\Delta t}{2}\Psi_{\text{c}}(t,\mathbf{x})\right]\psi_{\text{c}}(t, \mathbf{x}),
\end{align}
where $\mathcal{T}$ is the time ordering operator and, using the Baker–\allowbreak Campbell–\allowbreak Hausdorff formula, the time evolution operator has been split into three unitary parts which do not mix functions of the position and derivative operators. This makes it natural to automatically couple the method to particle-based $N$-body techniques that evolve collisionless components such as CDM and stellar particles on the same sub-time step spacing. Coupling to gas cells is also straightforward and is achieved via the full gravitational potential $\Psi_{\text{c}}$ (including the baryonic contribution) in both the SP equations and the forces evolving the gas cells. Simulations involving mixed ultralight and baryonic physics are left for future work.

The fields $\psi_{\text{c}}$ and $\Psi_{\text{c}}$ are discretised on a uniform Cartesian grid with $N^3$ mesh points in a periodic box of length $L$ to allow for efficient numerical computations using the Fast Fourier Transform (FFT). The pseudo-spectral method is summarised in Algorithm \ref{a_pseudospectral}.
\begin{algorithm}[t]
\caption{Pseudo-Spectral Method for MDM}\label{a_pseudospectral}
\begin{algorithmic}
\Require $\psi_{\text{c}}(t, \mathbf{x})$, $\Psi_{\text{c}}(t, \mathbf{x})$
\State $\psi_{\text{c}} \gets \exp\left[-i\frac{m}{\hbar}\frac{1}{a(t)}\frac{\Delta t}{2}\Psi_{\text{c}}(t,\mathbf{x})\right]\psi_{\text{c}}$ \Comment{kick}
\State $\psi_{\text{c}} \gets \mathrm{FFT}^{-1}\left(\exp\left[\frac{\hbar}{m}\frac{1}{a(t)^2}\frac{\Delta t}{2}k^2\right]\mathrm{FFT}(\psi_{\text{c}})\right)$ \Comment{drift}
\State $\Psi_{\text{c}} \gets \mathrm{FFT}^{-1}\left(-\frac{1}{k^2}\mathrm{FFT}\left(4\pi Gm(\rho_{\text{CDM,c}}+\rho_{\text{FDM,c}}-\bar{\rho}_{\text{m,c}})\right)\right)$ \Comment{update}
\State $\psi_{\text{c}} \gets \exp\left[-i\frac{m}{\hbar}\frac{1}{a(t)}\frac{\Delta t}{2}\Psi_{\text{c}}(t,\mathbf{x})\right]\psi_{\text{c}}$ \Comment{kick}
\State return $\psi_{\text{c}}(t+\Delta t, \mathbf{x})$, $\Psi_{\text{c}}(t+\Delta t, \mathbf{x})$
\end{algorithmic}
\end{algorithm}
The choice of the time step $\Delta t$ is determined by the requirement that the phase difference in the exponentials must not exceed $2\pi$, at which point the time step would be incorrectly `aliased' to a smaller time step corresponding to the phase difference subtracted by a multiple of $2\pi$ due to the periodicity of the exponential function. The kicks and the drift yield separate constraints for $\Delta t$, both of which must be simultaneously fulfilled. The resulting time step criterion is
\begin{equation}
\Delta t < \min\left(\frac{4}{3\pi}\frac{m}{\hbar}a^2\Delta x^2, 2\pi \frac{\hbar}{m}a\frac{1}{|\Psi_{\text{c},\text{max}}|}\right),
\label{e_timestep}
\end{equation}
where $\Delta x = L/N$ is the spatial resolution and $\Psi_{\text{c},\text{max}}$ is the maximum value of the potential. Note that Eq.~\eqref{e_timestep} is essentially a Courant–\allowbreak Friedrichs–\allowbreak Lewy (CFL) condition. The dependence $\Delta t \propto \Delta x^2$ can be viewed as a reflection of the relation of the Schrödinger equation to diffusion problems. Since $N$-body codes for gravity and Eulerian fluid solvers scale as $\propto \Delta x$, this adds computational cost to the simulations. However, Algorithm \eqref{a_pseudospectral} allows for machine precision control of the total kinetic energy and achieves spectral (exponential) convergence in space.

When considering the velocity field $\mathbf{v}_{\text{c}} = \mathbf{v}_{\text{M}} = \hbar \nabla_{\text{c}}\gamma/m$, another constraint on the validity of the discretisation becomes apparent. Since the difference in the gradient of the phase between two points can be at most $2\pi$, it follows that the discretised velocity field cannot exceed a maximum value (depending on the concrete form of the discretised gradient operator) of about
\begin{equation}
v_{\text{max}}=\frac{\hbar}{m}\frac{2\pi}{\Delta x}.
\label{e_vmax_eq}
\end{equation}
Velocities $v\geq v_{\text{max}}$ cannot be represented in a simulation with resolution $\Delta x$, which translates into a constraint on resolution, which should be good enough to resolve the de Broglie wavelength $\lambda_{\text{dB}}$ of the largest velocities:
\begin{equation}
\Delta x < \frac{\pi\hbar}{mv_{\text{max}}} \equiv \frac{1}{2}\lambda_{\text{dB}}(v_{\text{max}}).
\label{e_delta_x_req}
\end{equation}
Requirements \eqref{e_timestep} and \eqref{e_delta_x_req} exemplify why FDM simulations are computationally much more costly than traditional particle-based CDM simulations, where resolution can be set independent of velocities and time step constraints are less restrictive \citep[e.\,g.][]{Springel_2005_2}.

\subsection{Simulation Setup}
\label{ss_sim_setup_mdm}
As mentioned, in the MDM model only one of the DM particle species – occupying a fraction $f$ of the total matter content – is ultralight while all others are assumed to have negligible de Broglie wavelengths and are thus modelled as CDM. \cite{Schwabe_2020} pointed out that a soliton may not form in cosmological simulations with $f<0.1$. This result was confirmed by \cite{Lague_2023} who investigated the impact of a mixture of CDM and FDM in various proportions $f=[0, 1, 10, 50, 100] \,\%$ and for ultralight particle masses ranging over five orders of magnitude ($2.5 \times 10^{-25}$\,eV – $2.5 \times 10^{-21}$\,eV) using \textsc{AxioNyx}, albeit mostly for relatively small box sizes, $L_{\text{box}}=1$\,cMpc$/h$. The authors also implemented a modified friends-of-friends (FOF) halo finder and found good agreement between the inferred halo abundance and the predictions from the adapted halo model \textsc{AxionHMcode} in a narrow mass range. Expanding upon their findings, we aim to identify distinctive characteristics that set MDM apart from single-particle models, thereby enhancing our understanding of its cosmological implications.

\begin{table}
    \caption[Overview of the MDM simulation suite. All simulations are DM-only and assume an axion mass of $m = 10^{-24.5}\,\text{eV} = 3.16\times 10^{-25}\,\text{eV}$.]{Overview of the MDM simulation suite: (1)~FDM fraction $f$; (2)~side length of the simulation box $L_{\text{box}}$; (3)~number of CDM $N$-body particles; (4)~number of FDM grid cells; (5)~mass per CDM $N$-body particle. The softening scale for CDM is fixed to $\epsilon = 1.78$\,ckpc$/h$ in comoving units and capped at $\epsilon = 0.89$\,pkpc$/h$ in physical units following \cite{Power_2003}. All simulations are DM-only and assume an axion mass of $m = 10^{-24.5}\,\text{eV} = 3.16\times 10^{-25}\,\text{eV}$.}
    \label{t_sim_overview_mdm}

	\centering
    \begin{tabular}{c c c c c}
     \toprule
     $f$ & $L_{\text{box}}$ (cMpc/$h$) & $N_{\text{CDM}}$ & $N_{\text{FDM}}$ & $m_{\text{CDM}}$ ($10^7 \, M_{\odot}/h$) \\
     \midrule
     0.0\phantom{0} & $60$ & $1024^3$ & \rlap{NA}\phantom{$2048^3$} & $1.72$\\
     0.01 & $60$ & $1024^3$ & $2048^3$ & $1.71$\\
     0.1\phantom{0} & $60$ & $1024^3$ & $2048^3$ & $1.55$\\
     0.2\phantom{0} & $60$ & $1024^3$ & $2048^3$ & $1.38$\\
     0.3\phantom{0} & $60$ & $1024^3$ & $2048^3$ & $1.21$\\
     \bottomrule
    \end{tabular}
\end{table}

\begin{figure*}
\includegraphics[width=0.9\textwidth]{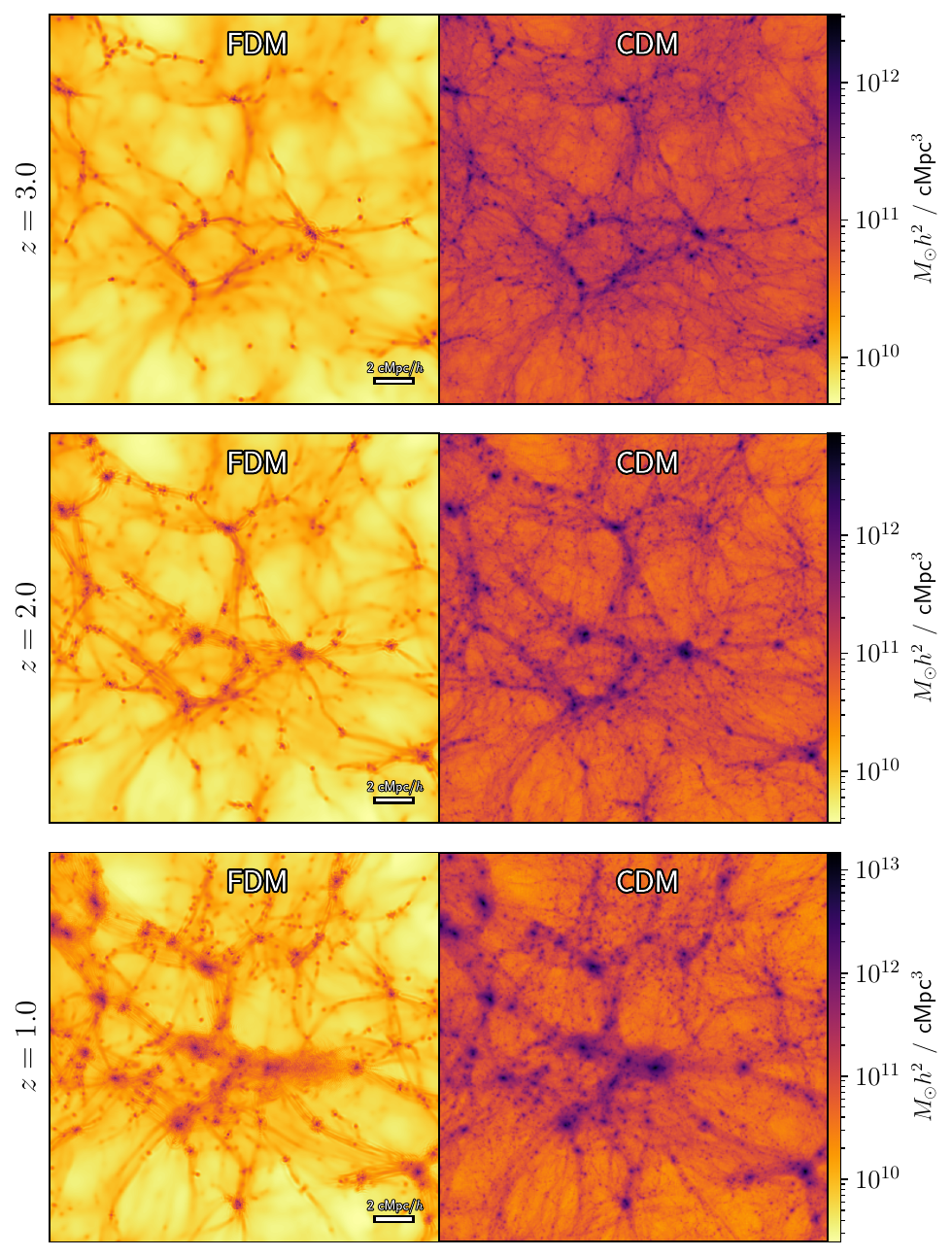}
\caption[MDM structure formation for $m=3.16\times 10^{-25}$\,eV and $f=0.1$. The redshift spacings (different rows) correspond to equal logarithmic spacings in the scale factor.]{MDM structure formation for $m=3.16\times 10^{-25}$\,eV and $f=0.1$. We plot projected (comoving) densities of the FDM (left column) and CDM (right column) components along the line of sight on a logarithmic scale. The redshift spacings (different rows) correspond to equal logarithmic spacings in the scale factor. The axion Jeans scale prevents FDM from clustering around small-mass halos, which thus remain CDM-dominated. The wave dynamics of FDM is reflected in interference fringes along filaments as well as the presence of granule structures in the DM halos, most visible towards lower redshift.}
\label{f_mdm_sim_projs}
\end{figure*}

For our MDM simulations, we adopt an axion mass of $m = 10^{-24.5} \,\text{eV} = 3.16\times 10^{-25} \,\text{eV}$ and vary the FDM fraction in the set $f\in [0.01, 0.1, 0.2, 0.3]$. Note that due to the presence of a dominant CDM component ($f < 0.5$), the potential wells in which the wavefunction evolves become steeper, which increases the FDM velocity dispersion and decreases its de Broglie wavelength $\lambda_{\text{dB}}$, Eq.~\eqref{e_delta_x_req}. MDM simulations thus require a higher resolution than their pure FDM ($f=f_{\text{max}}$) counterparts. Consequently, conducting large-scale cosmological simulations involving the solution of the full Schrödinger–Poisson system within the MDM framework is challenging. While we ascertain the necessary CDM and FDM resolution requirements according to the criteria outlined in Sec.~\ref{ss_pseudo} and Dome et al.\ (in prep), we thus ensure at the same time that our selection of simulation parameters is conservative. The final specifications of our MDM simulation suite are given in Table~\ref{t_sim_overview_mdm}.

To ensure the fidelity of the high redshift evolution of the simulations, we performed several tests: we verified that across several orders of magnitude the partial (FDM and CDM components) and total power spectra of the first snapshot ($z=127$) replicate the target power spectra obtained using the modified version of \textsc{AxionCamb} (see Sec.~\ref{ss_ics}); we also tested the linear growth prediction $D_{+}(k,a)\propto a$ (from linear theory) which holds on large scales for both partial and total power spectra. Finally, we exclude snapshots with redshifts below $z\lesssim 1$, as even with our conservative selection of simulation parameters, there is a risk that their highest velocity dispersions may remain unresolved by the FDM solver (see Eq.~\ref{e_delta_x_req}).

We visualise the MDM density distribution for an axion fraction $f=0.1$ in Fig.~\ref{f_mdm_sim_projs}, juxtaposing the FDM and CDM fields. The clustering of the former is visibly suppressed below the axion Jeans scale $k_{\text{J}}$, Eq.~\eqref{e_jeans_fdm}. The smallest-mass halos can remain completely CDM-dominated. As known from the pure FDM case ($f=f_{\text{max}}$), the wave dynamics are reflected in interference fringes along filaments as well as granule structures in the DM halos.

\section{Halo Mass Distribution and Density Profiles}
\label{s_mdm_stats}
In the following, we present a study of HMFs and halo density profiles in MDM cosmologies. To identify halos, we use the \textsc{Rockstar} halo finder \citep{Behroozi_2013}, which is based on adaptive hierarchical refinement of FOF groups in six dimensions (position and momentum space). This method provides robust tracking of substructure, being grid-independent, orientation-independent, and resilient to noise. However, \textsc{Rockstar} is a particle-based halo finder and thus can only be applied to particle distributions. Although it is theoretically possible to convert FDM grid cells into particles by concentrating the mass into points at the centre of each cell, we found\footnote{We modified \textsc{Rockstar-Galaxies}, an extended version of \textsc{Rockstar} with multi-mass and multi-type support, and applied it to the combined CDM and FDM `particles'. \textsc{Rockstar-Galaxies} is available at \url{https://bitbucket.org/pbehroozi/rockstar-galaxies/src/main/}.} that this approach leads to unreliable FOF groupings due to artifacts introduced by the grid features in the particle distribution.

Given that the dominant component in our MDM simulations is CDM ($f < 0.5$), we base our halo identification exclusively on the CDM component. This is achieved by running \textsc{Rockstar} on the (equal-mass) CDM particle distribution after correcting the CDM particle mass $m_{\text{CDM}}$ such that the total mass of CDM particles present within the simulation box aligns with the anticipated total DM mass.\footnote{Specifically, if $M_{\text{DM}} = \Omega_{\text{m}} \rho_{\text{crit}} L_{\text{box}}^3$ and $M_{\text{CDM}} = N_{\text{CDM}} m_{\text{CDM}}$ is the sum of the mass of all the CDM particles, then the CDM particle mass $m_{\text{CDM}}$ is upscaled by the factor $M_{\text{DM}} / M_{\text{CDM}}$.} We will justify the usage of \textsc{Rockstar} on the CDM component a posteriori.

\subsection{Halo Mass Functions}
\label{ss_hmfs}
\begin{figure*}
\includegraphics[width=\textwidth]{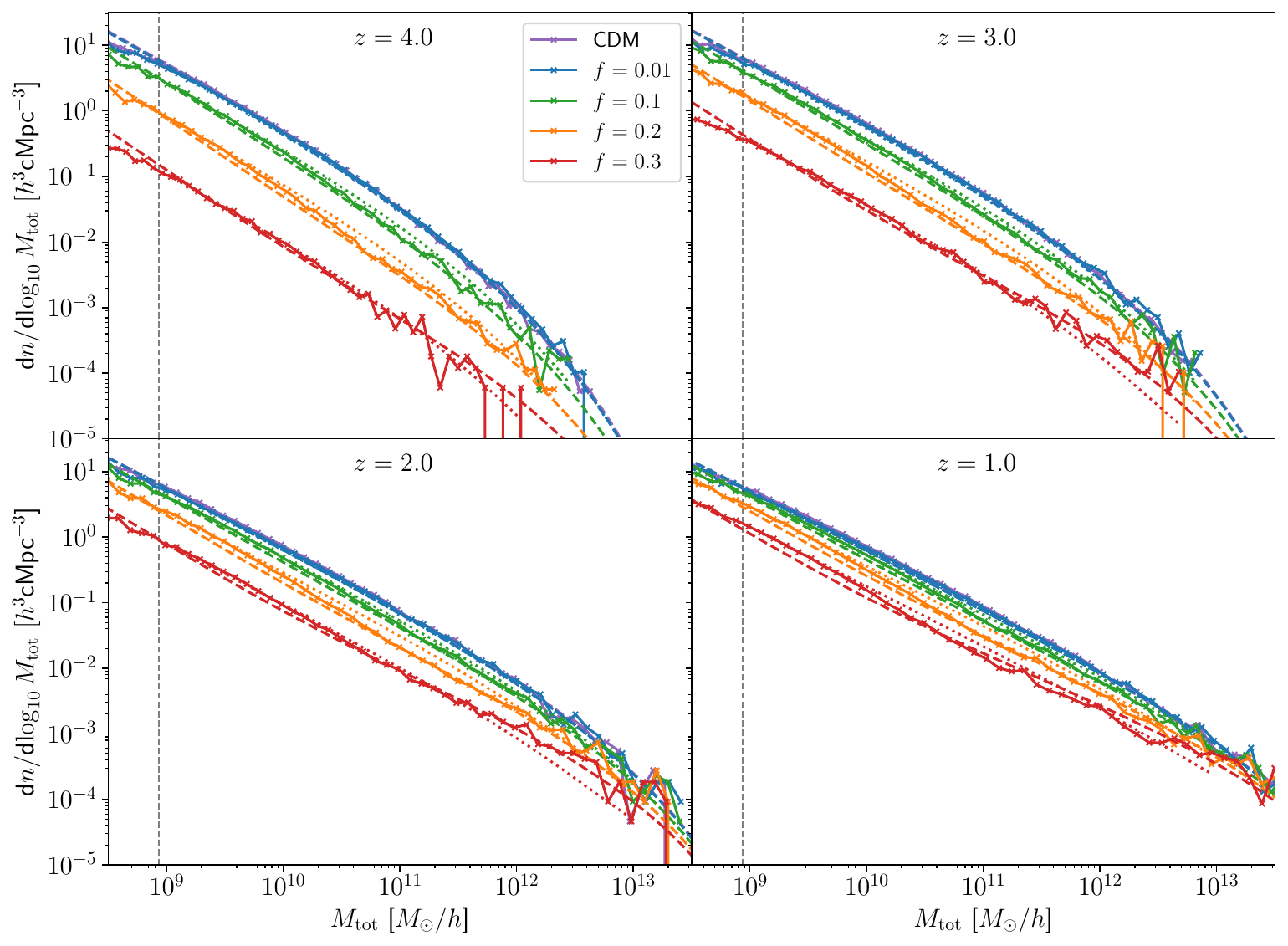}
\caption{MDM halo mass function (HMF) across redshifts $z=1-4$. We show results for a $m=3.16\times 10^{-25}$\,eV MDM cosmology at various FDM fractions $f$ (see legend) and the reference CDM cosmology. Solid lines with markers show the HMF inferred from the simulation. The dashed curves trace the Sheth–Tormen HMF (based on linear MDM power spectra), which show good agreement with the inferred HMF. Dotted curves are best-fit results against Eq.~\eqref{e_nhmf_model}, where we keep $\alpha_1=1.1$ and $\alpha_2=2.2$ fixed during the fit. The vertical dashed line corresponds to the mass resolution $50 \times m_{\text{CDM}}'$, where $m_{\text{CDM}}'$ is the rescaled CDM particle mass (see text).}
\label{f_mdm_hmf}
\end{figure*}

To measure the halo abundance in the simulations, we choose logarithmic mass bins of width $\Delta \log(M_{\mathrm{tot}}) = 0.03$ and count the number of \textsc{Rockstar} halos that fall into each bin. We show the resulting HMFs in Fig.~\ref{f_mdm_hmf} at redshifts $z=1-4$. While CDM and the $f=0.01$ cosmology follow the bottom-up structure formation paradigm in which small-mass halos form first and thus the small-mass end of their HMF barely changes from $z=4$ to $z=1$, MDM cosmologies with $f\gtrsim 0.1$ violate this picture and many small-mass halos are assembled at low redshift, with corresponding changes in the small-mass amplitude of their HMF.

We now compare against theoretical predictions for the HMF by employing the Sheth–Tormen approach for the halo mass function \citep{Press_1974, Sheth_2002},
\begin{equation}
\frac{1}{M}\frac{\mathrm{d}n}{\mathrm{d}\ln(M)} = \frac{1}{2}\frac{\bar{\rho}(z)}{M^2}f(\mathrm{\nu})\biggr\lvert\frac{\mathrm{d}\ln(\sigma^2)}{\mathrm{d}\ln(M)}\biggr\rvert,
\end{equation}
where $n$ is the halo number density, $\nu=\delta_{\text{crit}}/\sigma(M,z)$ is the peak height with $\delta_{\text{crit}}=1.686$ the critical linear density threshold for halo collapse, and the multiplicity function (for ellipsoidal collapse) is given by
\begin{equation}
f_{\text{ST}}(\nu)=A\sqrt{\frac{2}{\pi}}\sqrt{q}\nu(1+(\sqrt{q}\nu)^{-2p})e^{-\frac{q\nu^2}{2}},
\end{equation}
with $A=0.3222$, $p=0.3$ and $q=0.707$. We adopt a spherical top hat window function $W$ in real space when calculating the variance of the linear MDM power spectrum,\footnote{Note that when calculating the matter variance, \textsc{AxionHMcode} uses the linear power spectrum of cold matter, rather than the total matter linear power spectrum (see Sec.~\ref{s_calibrate}).}
\begin{align}
\sigma(R,z)^2 &= \frac{1}{2\pi^2}\int_{0}^{\infty} P^{\text{L}}(k,z)\tilde{W}(Rk)^2k^2\mathrm{d}k, \\
\tilde{W}(x) &= \frac{3}{x^3}(\sin x - x \cos x).
\end{align}
The variance can be transformed into a function of the halo mass via $M=4\pi\bar{\rho}R^3/3$.

\begin{table}
    \caption{Best-fit values of $\beta$ in Eq.~\eqref{e_nhmf_model} for our MDM simulation suite across a range of redshifts $z=1-8$. Note that the dependence on $f$ is substantial at higher redshift $z\gtrsim 3$. For $f=0.01$ the exact value of $\beta$ is less relevant since only the combination $\beta f$ enters the parametrisation \eqref{e_nhmf_model}.}
    \label{t_beta_bf}
    
	\centering
    \begin{tabular}{L{1.0cm}L{1.0cm}L{1.0cm}L{1.0cm}L{1.0cm}}
     \toprule
     \multirow{2}{*}{$z$} & \multicolumn{4}{c}{$f$} \\ \cline{2-5}
        & $0.01$ & $0.1$ & $0.2$ & $0.3$ \\
     \midrule
     \multicolumn{1}{c|}{$8.0$} & $32.7$ & $8.7$ & $5.0$ & $3.3$ \\ 
     \multicolumn{1}{c|}{$7.0$} & $25.8$ & $8.1$ & $4.9$ & $3.3$ \\
     \multicolumn{1}{c|}{$6.0$} & $22.0$ & $7.4$ & $4.9$ & $3.3$ \\
     \multicolumn{1}{c|}{$5.0$} & $19.4$ & $6.2$ & $4.6$ & $3.3$ \\
     \multicolumn{1}{c|}{$4.0$} & $13.8$ & $5.1$ & $4.3$ & $3.3$ \\
     \multicolumn{1}{c|}{$3.0$} & $12.0$ & $3.9$ & $3.6$ & $3.1$ \\
     \multicolumn{1}{c|}{$2.0$} & $\phantom{0}5.8$ & $2.6$ & $2.8$ & $2.9$ \\
     \multicolumn{1}{c|}{$1.0$} & $\phantom{0}3.6$ & $1.9$ & $2.2$ & $2.4$ \\ \bottomrule
    \end{tabular}
\end{table}

\begin{figure*}
\includegraphics[width=\textwidth]{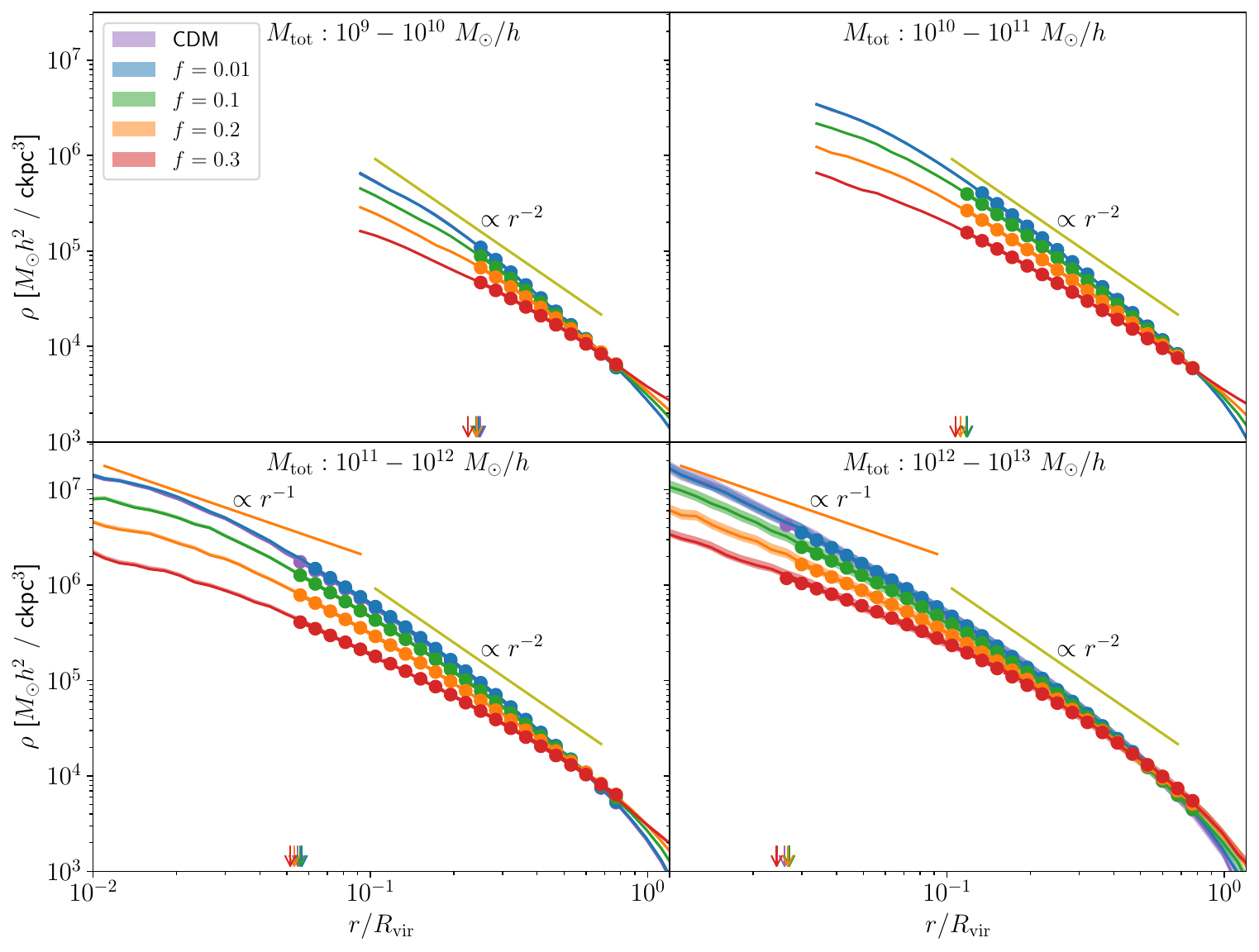}
\caption{Spherically averaged density profiles in MDM at redshift $z=1$. The mass bins range from $10^9 - 10^{10} \, M_{\odot}/h$ in the top left to $10^{12} - 10^{13} \, M_{\odot}/h$ in the bottom right. The shaded areas delineate the standard error of the median while the solid curves trace the median profile in each mass bin. Arrows at the bottom of the panels indicate inner convergence radii $r_{\text{conv}} = 5\epsilon$. The $r^{-1}$ line marks the characteristic inner slope of an NFW-like profile, while the $r^{-2}$ line serves as a reference for the transition radius used to define halo concentration. Einasto best-fits are shown by circular markers, and we find that three-parameter Einasto model provides a good parametrisation of MDM density profiles across the entire range of FDM fractions $f=0.0-0.3$.}
\label{f_rho_profs}
\end{figure*}

We show these model predictions as dashed curves in Fig.~\ref{f_mdm_hmf} and find a reasonable agreement between the model and the simulations, at all redshifts shown $z=1-4$ and in fact out to $z=10$ (not shown). The model curves for $f\gtrsim 0.1$ branch off from the CDM curves at the characteristic mass $M_0$ (see Eq.~\ref{e_M0}) below which the FDM HMF is suppressed with respect to the CDM case. While $M_0$ has been introduced for pure FDM ($f=f_{\text{max}}$), our MDM HMFs also branch off at approximately $M_0 = 2.34\times 10^{13} \, M_{\odot}/h$, suggesting that $M_0$ remains a useful and accurate mass scale even when $f < f_{\text{max}}$.

We also aim to provide an analytic form for the HMFs as a function of the FDM fraction $f$. To that end, we generalise the two-parameter model of \cite{Schive_2016} and write
\begin{equation}
n(M_{\text{tot}}) = n^{\text{ST}}(P_{\text{L}}^{\text{c}})\left[1-\beta f+\beta f\left(1+\left(\frac{M_{\text{tot}}}{M_0}\right)^{-\alpha_1}\right)^{-\alpha_2}\right],
\label{e_nhmf_model}
\end{equation}
where $n^{\text{ST}}(P_{\text{L}}^{\text{c}})$ is the Sheth–Tormen HMF in CDM. The steepness of the suppression is controlled by the parameter $\alpha_2$ (but also $\beta$ for $M$ significantly smaller than $M_0$) while the sharpness of the transition at $M \approx M_0$ is controlled by $\alpha_1$. Fitting this model to $N$-body simulations of pure FDM ($f=f_{\text{max}}$) results in $\alpha_1 = 1.1$, $\alpha_2 = 2.2$ \citep{Schive_2016}. \cite{Schive_2016} have taken special care when removing spurious halos and have found that the redshift-independent suppression $(1+(M_\text{h}/M_0)^{-1.1})^{-2.2}$ provides a good fit across a range of redshifts $z>4$. Subsequent works simulating bona fide FDM including the axion wave dynamics \citep{May_2022} have found that this parametrisation remains robust down to at least $z=3$. Our MDM HMFs are also well fit using these reference values for $\alpha_1$ and $\alpha_2$, hence we adopt them throughout this work rather than refitting them.

The only parameter that requires tuning is thus $\beta$. This parameter is crucial for achieving higher levels of suppression, $1-\beta f$, to the left of the characteristic mass $M_0$ than would be allowed by the value of $f$ alone. Best-fit values of $\beta$ are presented in Table~\ref{t_beta_bf}. Our results indicate that $\beta$ typically decreases as $f$ increases and as $z$ decreases. The dependence on $f$ is notably weaker at low redshift ($z \lesssim 3$). Specifically, the values $\beta = 3.1$, $\beta=2.9$ and $\beta=2.4$ provide a good fit at redshifts $z=3$, $z=2$ and $z=1$, respectively. The best-fit results at $z=1-4$ are depicted as dotted curves in Fig.~\ref{f_mdm_hmf}. The overall agreement with inferred HMFs is good across axion fractions $f=0.01-0.3$, indicating that the mass scale $M_0$ accurately characterises the suppression of the HMF despite being independent of $f$. The deviation from the inferred HMF at the high-mass end in case of $f=0.3$ reveals limitations of parametrisation \eqref{e_nhmf_model} as well as a potential dependence of $M_0$ on $f$.

\subsection{Density Profiles}
\label{ss_dens_profs}
A well-known prediction of pure FDM models ($f=f_{\text{max}}$) is the formation of solitonic cores at the centre of halos \citep{Schive_2014}. The situation is more complex in MDM, where in case of CDM-dominated models ($f<0.5$) we expect the soliton to be buried beneath the CDM central density cusp. Before we verify this intuitive picture, we first quantify the total DM density profile averages in various mass bins. To better sample the FDM field in the centre of halos, we interpolate FDM density values (which are defined on a Eulerian grid in the simulation code) trilinearly to inter-grid points, while CDM density profiles are obtained using simple \textsc{CosmicProfiles} routines \citep{Dome_2023_joss}. The density profiles are calculated around halo centres as identified by \textsc{Rockstar}, which neglects the FDM component. This approach is justified in our CDM-dominated models ($f<0.5$), where the CDM cusp dominates the gravitational potential in the centre and shapes its minimum. Note that the FDM core exhibits time-dependent behaviour with no stable centre, undergoing random walks by an amount of the order of the soliton radius on timescales $\tau \propto (m\sigma_{\text{FDM}}^2)^{-1}$, driven by perturbations away from the perfect stationary state \citep{Chiang_2021,Li_2021}.

The resulting density profiles in four mass bins from $M_{\mathrm{tot}} = 10^9-10^{13} \, M_{\odot}/h$ are shown in Fig.~\ref{f_rho_profs}. The solid curves show the median density
profile in logarithmic radial bins of width $\Delta \log(r) = 0.05$ while the shaded area delineates the standard error of the median. Note that in CDM the scale radius $r_s \equiv r_{-2}$ at which the logarithmic slope has the isothermal value of $-2$, i.\,e. $\mathrm{d} \ln(\rho) / \mathrm{d} \ln(r)|_{r_{-2}} = -2$, migrates towards larger normalised radii as the halo mass grows. In other words, the concentration 
\begin{equation}
c \equiv \frac{R_{\text{vir}}}{r_{-2}}
\end{equation}
of CDM halos decreases as the halo mass increases. This well-known result reflects the higher background density at earlier epochs when smaller mass halos form \citep{Navarro_1997, Bullock_2001, Ludlow_2014}.

To approximate the halo density profiles using an analytic form and to determine their concentration, we invoke the Einasto profile \citep{Einasto_1965}
\begin{equation}
\ln\left(\frac{\rho_{\text{E}}(r)}{\rho_{-2}}\right) = -\frac{2}{\zeta}\left[\left(\frac{r}{r_{-2}}\right)^{\zeta}-1\right]
\label{e_def_einasto}
\end{equation}
and assess the fidelity thereof a posteriori. Best-fit Einasto profiles are determined by adjusting the three parameters ($\zeta$, $r_{-2}$ and $\rho_{-2}$) of Eq.~\eqref{e_def_einasto} in order to minimise a figure-of-merit defined as
\begin{equation}
\psi^2 = \frac{1}{N_{\text{bin}}} \sum_{i = 1}^{N_{\text{bin}}} \left[\ln(\rho_i) - \ln(\rho_{\text{E}}(r_i; \rho_{-2}; r_{-2}; \zeta))\right]^2.
\end{equation}
We choose an inner convergence radius of $r_{\text{conv}} = 5\epsilon$, where $\epsilon$ is the gravitational softening length. This choice is based on our finding that the $3\epsilon$ recommendation by \cite{Power_2003} is not sufficiently conservative, particularly for small-mass bins, potentially leading to an underestimation of halo concentration. In addition, radial bins that exceed the outer limit of $0.8 \ R_{\text{vir}}$ are also discarded in the fit since they might correspond to radii where halos are not fully relaxed \citep{Ludlow_2016}. Best-fit results are shown in Fig.~\ref{f_rho_profs} and reproduce MDM density profiles very well across the entire range of FDM fractions $f=0.0-0.3$ at the redshift shown ($z=1$) but in fact out to $z=10$ (not shown), albeit at much lower statistical significance.

\subsection{Concentration–Mass Relation}
\label{ss_cM_relation}
\begin{figure}
\hspace{-.6cm}
\includegraphics[width=0.53\textwidth]{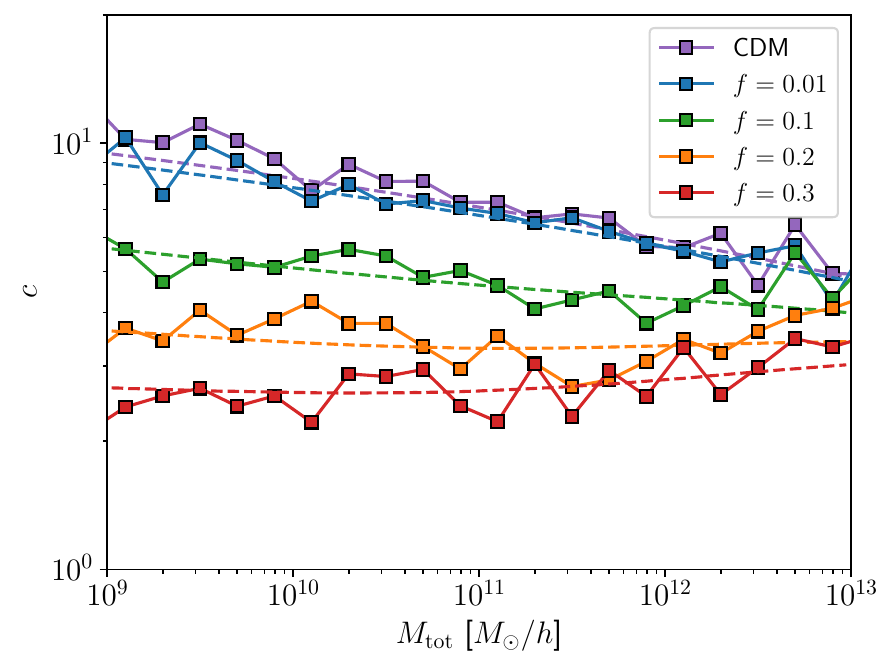}
\caption{Concentration–mass relation $c(M, z)$ in MDM cosmologies at redshift $z=1$. Square markers indicate median concentrations obtained from Einasto best-fits. Analytic model predictions from \protect\cite{Ludlow_2016} for parameters $f_{\text{coll}} = 0.02$ and $C = 650$ (see text) are traced by dashed curves, showing good agreement across $f=0.0-0.3$.}
\label{f_cM}
\end{figure}

We now turn to the concentration–mass relation $c(M,z)$ which follows trivially from the Einasto fits. The MDM concentration–mass relation inferred from the simulations at $z=1$ is shown in Fig.~\ref{f_cM} for logarithmic mass bins of width $\Delta \log(M_{\mathrm{tot}}) = 0.2$. While the CDM concentration follows the characteristic decrease toward higher mass as mentioned before, this decrease is weaker at higher FDM fractions $f$ and even seems to reverse for $f=0.3$. To model the concentration mass relation, we invoke the \cite{Ludlow_2016} approach based on extended Press–Schechter (EPS) theory. It stipulates that in order to estimate the mean concentration of halos at a given redshift $z_0$, one needs to solve the following system of coupled non-linear equations,
\begin{subequations}
\label{e_l16_model}
\begin{align}
\frac{\langle \rho_{-2}\rangle}{\rho_0} &= C\left(\frac{H(z_{-2})}{H(z_0)}\right)^2,\\
\frac{M_{-2}}{M_0} &= \text{erfc}\left(\frac{\delta_{\text{sc}}(z_{-2})-\delta_{\text{sc}}(z_0)}{\sqrt{2(\sigma^2(f_{\text{coll}}M_0)-\sigma^2(M_0))}}\right).
\end{align}
\end{subequations}
The first equation relates the mean inner density within the scale radius, $\langle \rho_{-2}\rangle$, with the critical density of the Universe at the collapse redshift, $z_{-2}$. The second equation expresses the collapse mass fraction using EPS theory \citep{Lacey_1993}, encapsulating the physical meaning of the collapse redshift as the redshift at which the characteristic mass, $M_{-2}$, was first contained in progenitors more massive than a fraction $f_{\text{coll}}$ of the final halo mass $M_0$.

Following the recommendations of \cite{Ludlow_2016}, we set $f_{\text{coll}} = 0.02$, $C=650$ and solve Eqs. \ref{e_l16_model} in the respective MDM cosmologies. Note that the linear variance $\sigma^2$ entering the equations depends on the linear MDM matter power spectrum. We show the analytic model predictions in Fig.~\ref{f_cM} and find good agreement with the inferred $c(M,z=1)$ relation across the entire range of FDM fractions $f=0.0-0.3$.

According to the model, the concentration $c$ exhibits a very flat $U$-shaped profile as a function of mass $M_{\text{tot}}$ for cosmologies with high values of $f$. It indicates that the strict bottom-up structure formation picture in which smaller-mass halos have necessarily higher concentration is invalidated in MDM cosmologies. Note that this result is well-known in pure FDM and warm DM (WDM) cosmologies \citep{Schneider_2012, Dome_2022}. Even though we cannot directly resolve the concentration mass scale $M_{\text{conc}}$ discussed around Fig.~\ref{f_4scales}, a visual extrapolation of the inferred concentration–mass relation $c(M,z)$ to higher masses suggests that the curves converge around $M_{\text{conc}}\approx 4.7 \times 10^{14} \, M_{\odot}/h$. This implies that $M_{\text{conc}} \approx 20-50 \, M_0$ represents a characteristic scale, independent of the axion fraction, at which the $c(M,z)$ relation in MDM diverges from that in CDM.

\section{Calibrating AxionHMcode}
\label{s_calibrate}
\subsection{Halo Model Details}
Having established that identifying halos based solely on the CDM component produces reliable HMFs and $c(M,z)$ relations, we now focus on calibrating the halo model framework \textsc{AxionHMcode} \citep{Vogt_2023} to the MDM simulations and compare the resulting non-linear power spectra. \textsc{AxionHMcode}, an adaptation of \textsc{HMCode-2020} \citep{Mead_2021}, incorporates ultralight particles as part of the DM. It builds on the halo model originally implemented in \cite{Dentler_2022} to constrain axions using weak lensing shear statistics. The code takes as input a range of cosmological and DM parameters: $\Omega_{\text{m}}$, $\Omega_{\text{b}}$, $f$, $m$, $H_0$, $n_s$, $A_s$, $k_{\text{piv}}$. Here, the axion density parameter is $\Omega_{\text{a}} \equiv \Omega_{\text{FDM}} = f\Omega_{\text{m}}$, while the combined density of CDM and baryons (`cold' matter) is $\Omega_{\text{c}} = \Omega_{\text{m}} - \Omega_{\text{a}} = \Omega_{\text{CDM}} + \Omega_{\text{b}}$. Using this input, \textsc{AxionHMcode} calculates the non-linear power spectrum at a desired redshift $z$. The linear FDM power spectrum, which is integral to the model, is calculated using the public version of \textsc{AxionCamb}. The total matter overdensity in the MDM cosmology can be decomposed into a sum of the cold matter, $\delta_{\text{c}}$, and axions, $\delta_{\text{a}}$,
\begin{equation}
\delta_{\text{m}} = \frac{\Omega_{\text{c}}}{\Omega_{\text{m}}}\delta_{\text{c}} + \frac{\Omega_{\text{a}}}{\Omega_{\text{m}}}\delta_{\text{a}}.
\end{equation}
The non-linear power spectrum in \textsc{AxionHMcode} is constructed as
\begin{equation}
P(k) = \left(\frac{\Omega_{\text{c}}}{\Omega_{\text{m}}}\right)^2P_{\text{c}}(k)+\frac{2\Omega_{\text{c}}\Omega_{\text{a}}}{\Omega_{\text{m}}^2}P_{\text{c,a}}(k)+\left(\frac{\Omega_{\text{a}}}{\Omega_{\text{m}}}\right)^2P_{\text{a}}(k),
\label{e_three_terms}
\end{equation}
where $P_{\text{c}}$, $P_{\text{c,a}}(k) \propto \delta_{\text{c}}\delta_{a}$ and $P_{\text{a}}$ are the cold, cross and axion power spectrum, respectively. For $P_{\text{c}}(k)$, the \textit{standard} halo model (see Sec.~\ref{ss_axionhmcode_params} for improvements on top of this) is adopted, splitting the power spectrum into the one-halo term $P^{1\text{h}}$ and two-halo term $P^{2\text{h}}$,
\begin{equation}
P_{\text{c}}(k) = P^{1\text{h}}_{\text{c}}(k)+P^{2\text{h}}_{\text{c}}(k).
\label{e_standard_hm}
\end{equation}
For axions we adopt the biased tracer formalism \citep{Massara_2014} which assumes that a sub-component, $\delta_{\text{L}}$, cannot cluster and evolves approximately linearly while the remaining fraction
\begin{equation}
F_{\text{h}} = \frac{1}{\bar{\rho}_{\text{a}}}\int_{M_{\text{cut}}}^{\infty}\mathrm{d}M_{\text{c}}n(M_{\text{c}})b(M_{\text{c}})M_{\text{a}}(M_{\text{c}})  \in [0,1]
\label{e_clustered_fraction}
\end{equation}
is in halos, i.\,e.
\begin{equation}
\delta_{\text{a}} = F_{\text{h}}\delta_{\text{h}}+(1-F_{\text{h}})\delta_{\text{L}}.
\end{equation}
Here, $n(M_{\text{c}})$ denotes the cold HMF, and we have introduced the cold halo bias, $b(M_{\text{c}})$, as well as the axion halo mass to cold halo mass relation, $M_{\text{a}}(M_{\text{c}})$. The biased tracer formalism assumes that axion halos only form in and around cold matter halos and thus the halo mass function for axions is the same as for the cold field, $n(M_{\text{a}})\mathrm{d}M_{\text{a}} = n(M_{\text{c}})\mathrm{d}M_{\text{c}}$, and the linear axion halo bias corresponds to the cold halo
bias, $b(M_{\text{a}}) = b(M_{\text{c}})$, i.\,e. $M_{\text{a}}$ is itself a function of $M_{\text{c}}$, a relation we need to specify (see below). In total there are three new quantities we have to provide to complete the MDM halo model: the cut-off mass, $M_{\text{cut}}$, the axion halo mass relation, $M_{\text{a}}(M_{\text{c}})$, and the axion halo density profile $\rho_{\text{a}}(r,M_{\text{a}},z)$. For details see \cite{Vogt_2023}.

To construct the cold halo density profile, we retain the \cite{Bullock_2001}-inspired model for the concentration parameter of cold halos of mass $M_{\text{c}}$,
\begin{equation}
c_{\text{c}}(M_{\text{c}},z) = B\left(\frac{1+z_{\text{f}}(M_{\text{c}},z)}{1+z}\right),
\label{e_Bullock2001}
\end{equation}
where $z_{\text{f}}$ denotes the formation redshift, which is defined by
\begin{equation}
\frac{D_{\text{+}}(z_{\text{f}})}{D_{\text{+}}(z)}\sigma_{\text{c}}(0.01 M_{\text{c}},z)=\delta_{\text{crit}},
\label{e_zf}
\end{equation}
with $D_{\text{+}}(z)$ denoting the total MDM linear growth rate (depending only on cosmology and redshift). The minimum halo concentration $B = 5.196$ is attained when the solution to Eq.~\eqref{e_zf} yields $z_{\text{f}} < z$. Even though the \cite{Bullock_2001} model has been shown to predict a sharp decline in the concentration at high mass that is inconsistent with $N$-body simulations and to insufficiently capture effects in non-CDM cosmologies such as WDM, the more accurate \cite{Ludlow_2016} approach introduced in Sec.~\ref{ss_dens_profs} is less suitable for the halo model since it typically underpredicts CDM small-scale power around $k\gtrsim 5 \, \text{cMpc}^{-1}$ in both pure CDM and MDM simulations. Shifting to \cite{Ludlow_2016} would also necessitate recalibrating \textsc{HMCode-2020} parameters (most notably the halo bloating parameter $\eta$) without gaining in precision.

\begin{figure}
\hspace*{-0.4cm}
\includegraphics[width=0.5\textwidth]{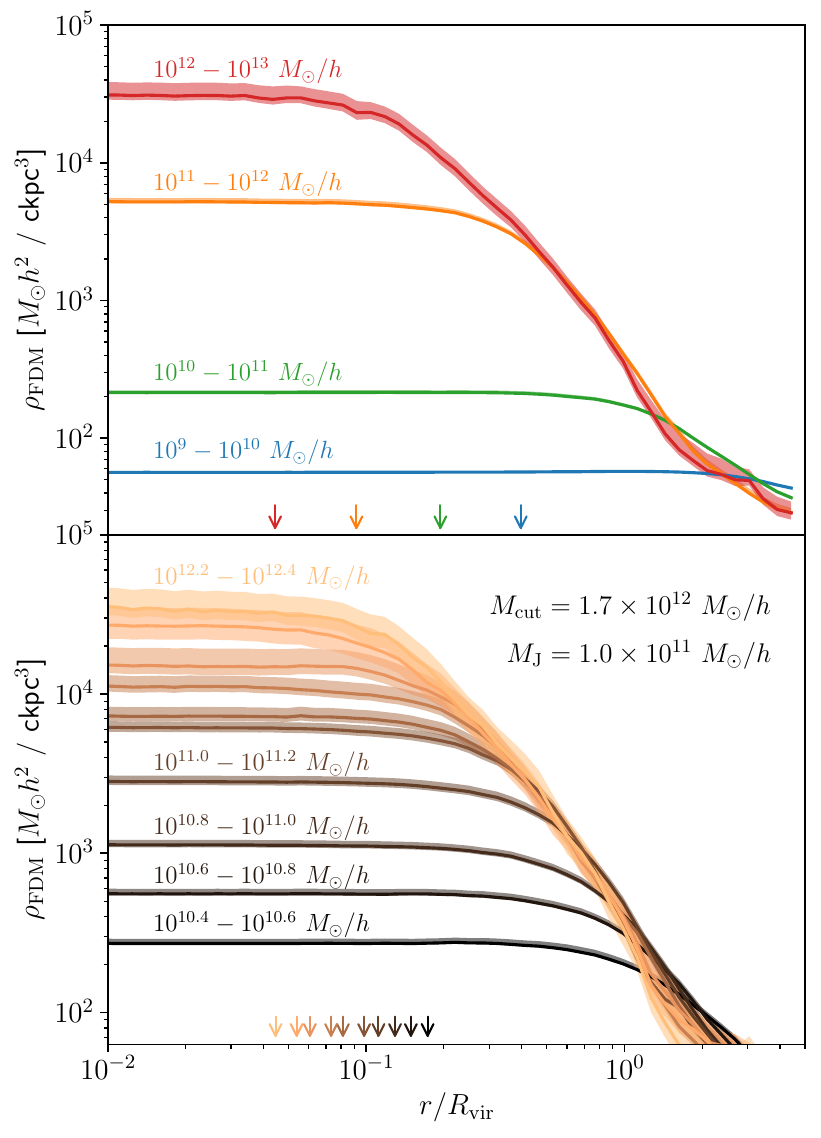}
\caption{Spherically averaged density profile of the FDM component in the $f=0.1$ MDM cosmology at redshift $z=1$. The top panel covers a total halo mass range of $M_{\text{tot}} = 10^9 - 10^{13} \, M_{\odot}/h$ while the lower panel focuses on the range $M_{\text{tot}}= 10^{10.4} - 10^{12.4} \, M_{\odot}/h$ with a finer mass resolution of $\Delta \log(M_{\text{tot}}) = 0.2$. Shaded areas delineate the standard error of the median, and solid curves trace the median profile in each mass bin. Arrows at the bottom of the panels indicate the FDM grid resolution scale of $\Delta x/2 = 14.6 \, \text{ckpc}/h$, rescaled by the mean virial radius in each mass bin. For masses $M_{\text{tot}} > M_{\mathrm{cut}} = 1.7 \times 10^{12} \, M_{\odot}/h$, the axion profile displays a distinct core-like feature, indicating the presence of a soliton. The steep decline in central densities between neighboring mass bins becomes pronounced for $M_{\text{tot}} < M_{\mathrm{J}} = 1.0\times 10^{11} \, M_{\odot}/h$, reflecting a diminished influence of the FDM component on structure formation.}
\label{f_Mcut}
\end{figure} 

\begin{figure*}
\includegraphics[width=\textwidth]{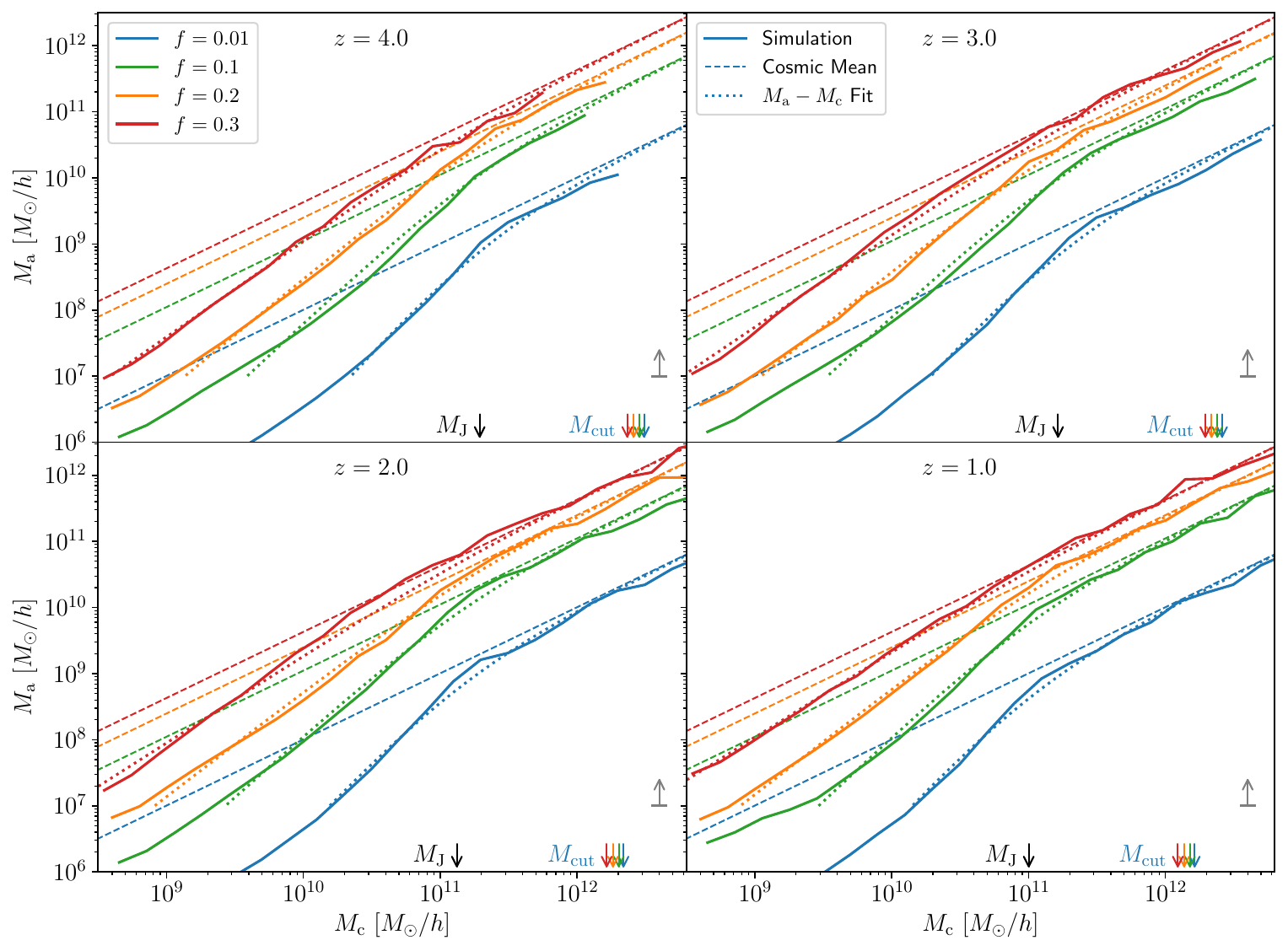}
\caption{Axion halo mass–cold halo mass relation $M_{\text{a}}(M_{\text{c}})$ in MDM cosmologies at various axion fractions $f=0.0-0.3$ across redshifts $z=1-4$. The mass resolution is $\Delta \log(M_{\text{tot}}) = 0.2$ as in Fig.~\ref{f_Mcut} and we estimate the axion halo mass $M_{\text{a}}=4\pi\int_{0}^{R_{\text{vir}}}\mathrm{d}r \, r^2\rho_{\text{FDM}}(r)$ by integrating the axion density profile out to the virial radius $R_{\text{vir}}$. For consistency with $M_{\text{a}}$, we estimate the cold halo mass $M_{\text{c}}$ likewise via integration (see text). Dashed curves indicate the cosmic mean relation $M_{\text{a}} = \frac{\Omega_{\text{a}}}{\Omega_{\text{c}}}M_{\text{c}}$, currently implemented in \textsc{AxionHMcode}. Colored arrows at the bottom of the panels indicate the cut-off mass $M_{\text{cut}} = 5 \times 10^{11}-10^{12} \, M_{\odot}/h$ while the black arrow denotes the Jeans mass $M_{\text{J}}$. Note that we infer a steeper $M_{\text{a}}(M_{\text{c}})$ relation than the cosmic average for $M_{\text{c}} \lesssim M_{\text{J}}$, with steepness increasing toward lower $f$. The transition range widens toward high axion fractions, covering $M_{\text{J}}-M_{\text{cut}}$ in some cases. We fit a broken power law, Eq.~\eqref{e_MaxMc}, for $M_{\text{a}}> 10^7 \, M_{\odot}/h$ (grey arrow) and show results as dotted curves.}
\label{f_MaMc}
\end{figure*} 

\subsection{Cut-Off Mass}
\label{ss_cutoff}
We first perform a sanity check of the cut-off mass $M_{\text{cut}}$ adopted by \textsc{AxionHMcode}. It is obtained by invoking the concept of the halo Jeans scale $r_{\text{hJ}}$ from Sec.~\ref{ss_4scales}. To verify the accuracy of Eq.~\eqref{e_rhalo_jeans_length} as the length scale below which axions fail to cluster inside DM halos, we calculate density profiles of the axion component in the MDM simulations. Fig.~\ref{f_Mcut} illustrates these profiles for the $f=0.1$ MDM cosmology at redshift $z=1$ across various mass bins. Our analysis, which includes simulations across axion fractions $f=0.0-0.3$ and redshifts up to $z=10$, consistently shows that halos with mass $M_{\text{tot}} \gtrsim M_{\text{cut}}$ exhibit a distinct soliton core. Below the cut-off mass, and particularly below the Jeans mass $M_{\text{J}}$, the central density of the core declines sharply, resulting in featureless axion profiles for halos with $M_{\text{tot}} \ll M_{\text{J}}$.

This steep decline in profile density supports $M_{\text{cut}}$ as an effective estimate of the mass scale below which halos (slowly) transition to being predominantly CDM-dominated in MDM. It is important to note that in the transitional mass range $M_{\text{J}}-M_{\text{cut}}$, `quantum pressure' continues to play a role but is insufficient to fully counteract gravitational effects (recall definition of $M_{\text{J}}$ in Sec.~\ref{ss_4scales}). The transition to CDM-dominated halos can occur over a large mass range exceeding 0.5 dex, and is not always complete for mass $M_{\text{tot}} \approx M_{\text{J}}$. In the following, we study integrated axion density profiles (i.\,e. the axion halo mass relation) and we will see that the Jeans mass and cut-off mass can help describe the transition toward CDM-dominated halos.

\subsection{Axion Halo Mass Relation}
\label{ss_axionhalomass}
We now turn to the relationship between the axion halo mass, $M_{\text{a}}$, and the cold halo mass, $M_{\text{c}}$. In the public version of \textsc{AxionHMcode}, 
it is assumed that the axion mass $M_{\text{a}}$ follows the cosmic abundance fraction relative to the cold halo mass $M_{\text{c}}$ down to the cut-off mass $M_{\text{cut}}$, i.\,e. that $M_{\text{a}} = (\Omega_{\text{a}} / \Omega_{\text{c}}) M_{\text{c}}$ for $M_{\text{c}} > M_{\text{cut}}$. However, as seen in Sec.~\ref{ss_cutoff}, there are strong indications that this simplistic approach may not accurately capture the axion halo mass–cold halo mass relation found in simulations. Fig.~\ref{f_MaMc} shows the $M_{\text{a}}(M_{\text{c}})$ relation inferred from our MDM simulations at various axion fractions $f=0.0-0.3$ across redshifts $z=1-4$. We estimate both the axion and cold halo mass by integrating the respective density profile out to the virial radius $R_{\text{vir}}$,
\begin{equation}
M_{\text{a,c}}=4\pi\int_{0}^{R_{\text{vir}}}\mathrm{d}rr^2\rho_{\text{a,c}}(r).
\label{e_Mac_integration}
\end{equation}
The cold halo mass could also be obtained directly from the \textsc{Rockstar} virial mass, $M_{\text{c}} = (\Omega_{\text{c}}/\Omega_{\text{m}}) M_{\text{tot}}$. While generally consistent within $0.1$ dex, we find deviations between the two $M_{\text{c}}$ estimates of up to $0.5$ dex at the high-mass end, particularly at high redshift. These discrepancies arise from low halo number statistics (see e.\,g. Fig.~\ref{f_mdm_hmf}), which lead to larger uncertainties in the median density profile $\rho_{\text{c}}$ for high-mass bins. To ensure consistency with the calculation of $M_{\text{a}}$, we determine $M_{\text{c}}$ via integration following Eq.~\eqref{e_Mac_integration}.

\begin{table}
    \caption{Best-fit values of $\beta_2$ in Eq.~\eqref{e_MaxMc} for our MDM simulation suite across a range of redshifts $z=1-8$. In the fitting process, we fix $\beta_1$ at $1$, as allowing $\beta_1$ to vary only results in minor improvements. Note that $\beta_2$ increases with higher redshift and lower values of $f$.}
    \label{t_MaxMc}

	\centering
    \begin{tabular}{L{1.0cm}L{1.0cm}L{1.0cm}L{1.0cm}L{1.0cm}}
     \toprule
     \multirow{2}{*}{$z$} & \multicolumn{4}{c}{$f$} \\ \cline{2-5}
        & $0.01$ & $0.1$ & $0.2$ & $0.3$ \\
     \midrule
     \multicolumn{1}{c|}{$8.0$} & $1.52$ & $1.04$ & $0.78$ & $0.62$ \\
     \multicolumn{1}{c|}{$7.0$} & $1.48$ & $1.04$ & $0.77$ & $0.63$ \\
     \multicolumn{1}{c|}{$6.0$} & $1.46$ & $1.02$ & $0.79$ & $0.58$ \\
     \multicolumn{1}{c|}{$5.0$} & $1.43$ & $1.04$ & $0.75$ & $0.51$ \\
     \multicolumn{1}{c|}{$4.0$} & $1.37$ & $0.96$ & $0.71$ & $0.45$ \\
     \multicolumn{1}{c|}{$3.0$} & $1.32$ & $0.93$ & $0.66$ & $0.40$ \\
     \multicolumn{1}{c|}{$2.0$} & $1.22$ & $0.87$ & $0.59$ & $0.32$ \\
     \multicolumn{1}{c|}{$1.0$} & $1.27$ & $0.98$ & $0.61$ & $0.30$ \\
     \bottomrule
    \end{tabular}
\end{table}

As shown in Fig.~\ref{f_MaMc}, the cosmic average relation, $M_{\text{a}} = (\Omega_{\text{a}}/\Omega_{\text{c}}) M_{\text{c}}$, is a valid approximation at the high-mass end.\footnote{Note that while the total halo cut-off mass $M_{\text{cut}}$ from Sec.~\ref{ss_4scales} is independent of $f$, the corresponding cold matter values highlighted in Fig.~\ref{f_MaMc} do have a dependence on $f$.} Toward lower mass, the simulated $M_{\text{a}}(M_{\text{c}})$ relation becomes significantly steeper. Deviations from the cosmic mean exceed $1$ dex in several mass bins. These deviations become particularly pronounced below the Jeans mass, $M_{\text{J}}$, consistent with the axion density profiles discussed in Sec.~\ref{ss_cutoff}. The break can span a wide mass range, especially at higher axion fractions, typically encompassing $M_{\text{J}}-M_{\text{cut}}$ and exceeding 0.5 dex for high values of $f$. To better capture this behaviour, we propose to parametrise the axion halo mass–cold halo mass relation as a broken power law:
\begin{equation}
M_{\text{a}} = \left(1+\left(\frac{M_{\text{c}}}{M_{\text{J}}}\right)^{-\beta_1}\right)^{-\beta_2}\frac{\Omega_{\text{a}}}{\Omega_{\text{c}}}M_{\text{c}},
\label{e_MaxMc}
\end{equation} 
where the steepness of the suppression is controlled by the parameter $\beta_2$ while the sharpness of the transition
at $M \approx M_{\text{J}}$ is controlled by $\beta_1$, similar to Eq.~\eqref{e_nhmf_model}. The linear Jeans mass is given by Eq.~\eqref{e_Jeansmass}. Letting $\beta_1$ vary during the fit along with $\beta_2$ leads to only minor improvements, hence we fix $\beta_1 = 1$. Best-fit values of $\beta_2$ for $M_{\text{a}} > 10^7 \, M_{\odot}/h$ are presented in Table~\ref{t_MaxMc}, and the corresponding best-fit curves are illustrated in Fig.~\ref{f_MaMc}. We observe a weak dependence of $\beta_2$ on redshift, but a notable decrease of $\beta_2$ as the axion fraction $f$ increases, indicating a weaker suppression. This analysis shows that MDM cosmologies dominated by CDM $f<0.5$ are different from pure axion cosmologies ($f=f_{\text{max}}$), where axion halos do not exist below $M_{\text{J}}$ and virialised axion halos only form above $M_{\text{cut}}$ (see Sec.~\ref{ss_4scales}).

Another implication of the $M_{\text{a}}(M_{\text{c}})$ relation deviating from the cosmic mean, where $M_{\text{a}} = (\Omega_{\text{a}}/\Omega_{\text{c}}) M_{\text{c}}$, is that expressing total DM density profiles as $\rho_{\text{DM}}(r) = f\rho_{\text{FDM}}(r)+(1-f)\rho_{\text{CDM}}(r)$ \citep[see e.\,g.][]{Shevchuk_2023} can at best be a good approximation at high masses above the cut-off mass $M_{\text{cut}}$. A more detailed analysis of axion density profiles, including their parametrisation via soliton+NFW profiles, their (non-)formation at low axion fractions \citep[see e.\,g.][]{Schwabe_2020}, and their use in axion forecasts and constraints, is beyond the scope of this paper and will be addressed in future work.

\subsection{\textsc{AxionHMcode} Parameters}
\label{ss_axionhmcode_params}
The updated version of \textsc{AxionHMcode} has several improvements over the original implementation by \cite{Vogt_2023}, which we now summarise. As in \cite{Vogt_2023}, we adopt the \textsc{HMCode-2020} parameters \citep{Mead_2021}, which have been introduced to improve the model in its fit to $\Lambda$CDM simulations over the standard halo model. The \textsc{HMCode-2020} parameters were calibrated using the \textsc{Mira Titan} matter power spectrum emulator of \cite{Heitmann_2016, Lawrence_2017}. This cosmic emulator encompasses eight cosmological parameters and provides an accuracy of 4\,\% for $k < 7 \,h\,$cMpc$^{-1}$. In turn, the accuracy of \textsc{HMCode-2020} when compared to simulated $\Lambda$CDM data is excellent with a RMS error of less than 2.5\,\% for $k < 10 \,h\,$cMpc$^{-1}$ and $z<2$ \citep{Mead_2021}.

The \textsc{HMCode-2020} parameters are only calibrated up to $z=2$ and it is a priori not guaranteed that a non-linear power spectrum with these parameters at $z>2$ is more accurate than without the parameters (i.\,e.\ standard halo model). We adopt the \textsc{HMCode-2020} parameters up to $z=3.5$ and will show in Sec.~\ref{ss_nonlinear} that not only is the agreement with simulations within the 10\,\% margin for pure $\Lambda$CDM, but that the improvements to \textsc{AxionHMcode} (including generalising and recalibrating one of the \textsc{HMCode-2020} parameters) yield good agreement with simulated non-linear power spectra up to at least $z=3.5$ for axion fractions that we can assess, $f<0.3$.

\begin{table*}
    \caption{\textsc{AxionHMcode} parameters, built on top of the standard halo model. For each parameter, we provide its description, the equation defining it, the default value in the standard halo model, the fitted functional form or value, and an example of this function evaluated at $z=1$ for a standard MDM cosmology with $f=0.1$, $m=3.16\times 10^{-25}$\,eV and \protect\cite{Planck_2015} cosmological parameters.}
    \label{t_AxionHMcode}

    \renewcommand{\arraystretch}{1.6}
    \centering
    \begin{tabular}{llccll}
    \toprule
    Parameter & Explanation & Equation & Default & Fitted functional form or value & Example \\
    \midrule
    $k_{\text{d}}$ & Two-halo term damping wavenumber & \eqref{e_p2h_axhmcode} & $0$ & $0.05699 \times \sigma_{8,\text{c}}(z)^{-1.089} \, h \, \text{cMpc}^{-1}$ & $0.127 \, h \, \text{cMpc}^{-1}$ \\ 
    $q$ & Two-halo term fractional damping & \eqref{e_p2h_axhmcode} & $0$ & $0.2696 \times \sigma_{8,\text{c}}(z)^{0.9403}$ & $0.135$ \\ 
    $n_{\text{d}}$ & Two-halo term damping power & \eqref{e_p2h_axhmcode} & $1$ & $2.853$ & $2.85$ \\ 
    $k_{\ast}$ & One-halo term damping wavenumber & \eqref{e_p1h_axhmcode} & $0$ & $0.05618 \times \sigma_{8,\text{c}}(z)^{-1.013} \, h \, \text{cMpc}^{-1}$ & $0.118 \, h \, \text{cMpc}^{-1}$ \\ 
    $\eta$ & Halo bloating & \eqref{e_halobloating} & $0$ & $0.1281 \times \sigma_{8,\text{c}}(z)^{-0.3644}$ & $0.168$ \\ 
    $B$ & Minimum halo concentration & \eqref{e_Bullock2001} & $4$ & $5.196$ & $5.20$ \\ 
    $\alpha_0$ & \textsc{HMCode-2020} smoothing & \eqref{e_meadsmoothing} & $1$ & $1.875\times (1.603)^{n_{\text{eff,c}}(z)}$ & $0.560$ \\
    $\alpha_{1,\text{c}}$ & Small-scale cold-cold smoothing & \eqref{e_alpha1} & $1$ & $\alpha_0\left(1+0.124\left(\frac{10^{-24}}{m}\right)^{0.0450}\left(\frac{\Omega_{\text{a}}}{\Omega_{\text{m}}}\right)^{0.226}(1+z)^{1.13}\right)^{-1}$ & $0.475$ \\ 
	$\alpha_{2,\text{c}}$ & Large-scale cold-cold smoothing & \eqref{e_alpha2} & 1 & $\max\left(\alpha_{1,\text{c}}, \frac{1.10}{1+z}\right) \ \text{for} \ f \geq 0.01$, else $\alpha_{1,\text{c}}$ & $0.530$ \\
	$\alpha_{\text{c,a}}$ & Cold-axion (cross) smoothing & \eqref{e_alpha1} & $1$ & $\alpha_0\left(1+0.0487\left(\frac{10^{-24}}{m}\right)^{0.0450}\left(\frac{\Omega_{\text{a}}}{\Omega_{\text{m}}}\right)^{0.224}(1+z)^{2.21}\right)^{-1}$ & $0.485$ \\
	$\beta_2$ & Steepness of $M_{\text{a}}(M_{\text{c}})$ relation & \eqref{e_MaxMc} & $0$ & refer to tabulated values in Table~\ref{t_MaxMc} & $0.972$ \\
     \bottomrule
    \end{tabular}
\end{table*}

One of the \textsc{HMCode-2020} parameters is the one-halo term damping. In the standard halo model approach (see Eq.~\ref{e_standard_hm}), the one-halo term is typically constant on large scales. However, this does not accurately reflect mass and momentum conservation. It was demonstrated by \cite{Smith_2003} that the one-halo term should increase as $P_{\text{1h}}(k)\propto k^4$ at small $k$ (i.\,e., it should dampen compared to a constant at small $k$). To address this, \textsc{HMCode-2020} implements a modification:
\begin{equation}
P^{\text{1h}}_{\text{c}}(k)\rightarrow P^{\text{1h}}_{\text{c}}(k)\frac{(k/k_{\ast})^4}{1+(k/k_{\ast})^4}.
\label{e_p1h_axhmcode}
\end{equation}
This adjustment ensures that the one-halo term grows as expected and is suppressed on large scales. Consequently, on large scales, the non-linear power spectrum is primarily determined by the two-halo term, which aligns with the (perturbed) linear power spectrum. The suppression effect is controlled by the free parameter $k_{\ast}$, which was fitted as:
\begin{equation}
k_{\ast} = 0.05618\times \sigma_{8,\text{c}}(z)^{-1.013} \, h \, \text{cMpc}^{-1}.
\end{equation}
We can now list the improvements made in the new version of \mbox{\textsc{AxionHMcode}}:
\begin{itemize}
\item To construct the two-halo term for the cold matter component, we apply a perturbative damping to the linear power spectrum, so that
\begin{equation}
P^{\text{2h}}_{\text{c}}(k) = P^{\text{L}}_{\text{c}}(k)\left(1-q\frac{(k/k_{\text{d}})^{n_{\text{d}}}}{1+(k/k_{\text{d}})^{n_{\text{d}}}}\right).
\label{e_p2h_axhmcode}
\end{equation}
This formulation aligns with the recommendations of \cite{Mead_2021}, who suggest that, given the precision of modern halo models, it is crucial to account for the largest-scale non-linear effects. Specifically, perturbation theory indicates that the most significant non-linear effect on large scales is a small damping of power. The term introduced in Eq.~\eqref{e_p2h_axhmcode} is designed to capture this effect, with the three parameters fitted and given by \cite{Mead_2021},
\begin{align}
k_{\text{d}} &= 0.05699\times \sigma_{8,\text{c}}(z)^{-1.089} \, h \, \text{cMpc}^{-1},\nonumber \\ 
q &= 0.2696\times \sigma_{8,\text{c}}(z)^{0.9403},\\
n_{\text{d}} &= 2.853.\nonumber
\end{align}
Note that the difference between perturbatively corrected linear theory and the standard two-halo term is tiny but significant gains in computational time are made when replacing the integral expression of the standard two-halo term by Eq.~\eqref{e_p2h_axhmcode}.
\item We achieve additional significant speed-ups through modularisation and in-memory storage of arrays. For example, the formation redshift $z_{\text{f}}$ for a given cold halo mass $M_{\text{c}}$ is computed once and subsequently stored. As a result, we reduce the runtime on a single-core machine to under $1$ minute for a single evaluation of \textsc{AxionHMcode}, with only a minor increase in memory overhead.
\item We fix a bug in the implementation of the halo bloating effect mediated by the parameter $\eta$ which scales the wavenumber $k$ in the Fourier transformation of the NFW profile of cold halos as
\begin{equation}
\tilde{u}(k,M,z) \rightarrow \tilde{u}(\nu^{\eta} k,M,z).
\label{e_halobloating}
\end{equation}
The halo bloating parameter was fitted to \citep{Mead_2021}
\begin{equation}
\eta = 0.1281\times \sigma_{8,\text{c}}(z)^{-0.3644},
\end{equation}
which we also adopt in \textsc{AxionHMcode}.
\item We adjust the cosmic axion halo mass–cold halo mass relation $M_{\text{a}}(M_{\text{c}}) = (\Omega_{\text{a}}/\Omega_{\text{c}}) M_{\text{c}}$ by incorporating a broken power law below the linear Jeans mass $M_{\text{J}}$, in accordance with the findings from our MDM simulations which reveal a strong decrement in the inferred axion halo mass $M_{\text{a}}$ compared to the cosmic mean relation (see Eq.~\ref{e_MaxMc}). Consequently, lower integration limits for various quantities, such as the clustered fraction of Eq.~\eqref{e_clustered_fraction}, are adjusted from $M_{\text{cut}}$ down to $0$.
\item We continue to use the smoothing parameter $\alpha$, which allows us to overcome the simplistic assumption of a purely additive behaviour of one- and two-halo terms by modelling
\begin{equation}
P_{\text{c}}(k)= (P^{\text{1h}}_{\text{c}}(k)^{\alpha}+P^{\text{2h}}_{\text{c}}(k)^{\alpha})^{1/\alpha}.
\label{e_smoothing_default}
\end{equation}
\cite{Mead_2021} fitted parameter $\alpha$ to a general form,
\begin{equation}
\alpha = 1.875\times (1.603)^{n_{\text{eff,c}}(z)},
\label{e_meadsmoothing}
\end{equation}
where the effective spectral index at the non-linear length scale is
\begin{equation}
n_{\text{eff,c}}(z) = -\frac{\mathrm{d}\ln(\sigma_{\text{c}}^2(R,z))}{\mathrm{d}\ln(R)}\biggr\rvert_{\sigma_{\text{c}}=\delta_{\text{crit}}} - 3.
\end{equation}
We first generalise the smoothing parameter to the cold matter component of MDM. To prevent strong smoothing effects to spill over from the quasi-linear regime into the large-scale regime, the cold-cold smoothing is modeled using a logistic function between $\alpha_{1,\text{c}}$ and $\alpha_{2,\text{c}}$:
\begin{equation}
\alpha_{\text{c}}(k,z) = \alpha_{2,\text{c}}(z) + \frac{\alpha_{1,\text{c}}(z) - \alpha_{2,\text{c}}(z)}{1 + \exp((k_{\text{piv}} - k)/\Delta k)}.
\label{e_smoothing_coldcold}
\end{equation}
The width of the transition is controlled by $\Delta k = 0.1 \ h \ \text{cMpc}^{-1}$. In the small-scale limit, we use the following expression:
\begin{equation}
\alpha_{1,\text{c}}(z) = \frac{1.875\times (1.603)^{n_{\text{eff,c}}(z)}}{1+a\left(\frac{10^{-24}}{m}\right)^b\left(\frac{\Omega_{\text{a}}}{\Omega_{\text{m}}}\right)^d(1+z)^e}.
\label{e_alpha1}
\end{equation}
This parametrisation reduces to the default smoothing parameter for $\Lambda$CDM in the limit of a high axion mass $m \rightarrow \infty$ or a low axion fraction $\Omega_{\text{a}} \rightarrow 0$, given that $b,d >0$. Due to our lack of MDM simulation for axion mass values other than $m=3.16\times 10^{-25}$ eV, we fix the exponent $b=0.0450$ based on the expectation of a weak dependence on $m$. In the large-scale limit, where $k < k_{\text{piv}} = 0.5 \ h \ \text{cMpc}^{-1}$, the smoothing is well fit by
\begin{equation}
\alpha_{2,\text{c}}(z) = 
\begin{cases}
\alpha_{1,\text{c}}(z) \ \ &\text{for} \ f < 0.01 \\
\max\left(\alpha_{1,\text{c}}(z), \frac{1.10}{1+z}\right) \ \ &\text{for} \ f \geq 0.01.
\end{cases}
\label{e_alpha2}
\end{equation}
We fit parameters $a,d,e$ against the simulated cold-cold power spectrum $P_{\text{c}}$ (see Eq.~\ref{e_smoothing_default}) using the \cite{Nelder_1965} simplex algorithm with equal logarithmic weights in $k \in [0.2 - 30] \ h \ \text{cMpc}^{-1}$ and using data from $z = 1, 2, 3$ and $f = 0.0, 0.01, 0.1, 0.2, 0.3$ with equal weight, minimising the figure-of-merit
\begin{equation}
\psi^2 = \frac{1}{N}\sum_{z, f, k}|\log(\text{AxHM}(k,z,f))-\log(\text{Sim}(k,z,f))|^2.
\end{equation}
In addition, we find better agreement with simulations if we add an independent smoothing parameter to the cold-axion cross power spectrum $P_{\text{c,a}}$ (which depends on $P_{\text{c}}$):
\begin{equation}
P_{\text{c,a}}(k)=F_{\text{h}}\left(P^{\text{1h}}_{\text{c,a}}(k)^{\alpha_{\text{c,a}}}+P^{\text{2h}}_{\text{c,a}}(k)^{\alpha_{\text{c,a}}}\right)^{1/\alpha_{\text{c,a}}} + (1-F_{\text{h}})\sqrt{P_{\text{c}}(k)P^{\text{L}}_{\text{a}}(k)}
\end{equation}
We parametrise $\alpha_{\text{c,a}}$ in the exact same way as the small-scale cold-cold smoothing term $\alpha_{1,\text{c}}(z)$ in Eq.~\eqref{e_alpha1}, and fit parameters $a,d,e$ against the simulated $P_{\text{c,a}}$ spectrum. Best-fit values for $\alpha_{1,\text{c}}, \alpha_{2,\text{c}}$ and $\alpha_{\text{c,a}}$, along with all other \mbox{\scshape{AxionHMcode}} \normalfont parameters, are provided in Table~\ref{t_AxionHMcode}.
\item In the new version of \textsc{AxionHMcode}, we activate the halo bloating parameter $\eta$ (for the cold halo density profile), along with the one-halo damping (see Eq.~\ref{e_p1h_axhmcode}), the two-halo damping (see Eq.~\ref{e_p2h_axhmcode}), and the smoothing parameter (see Eqs.~\ref{e_alpha1} and \ref{e_smoothing_coldcold}) by default. This configuration significantly enhances the agreement with MDM simulations, both qualitatively and quantitatively. These improvements validate the necessity of \textsc{HMCode-2020}-like parameters for accurate MDM simulations, as initially suggested by \cite{Vogt_2023}.
\end{itemize}

\begin{figure}
\includegraphics[width=0.5\textwidth]{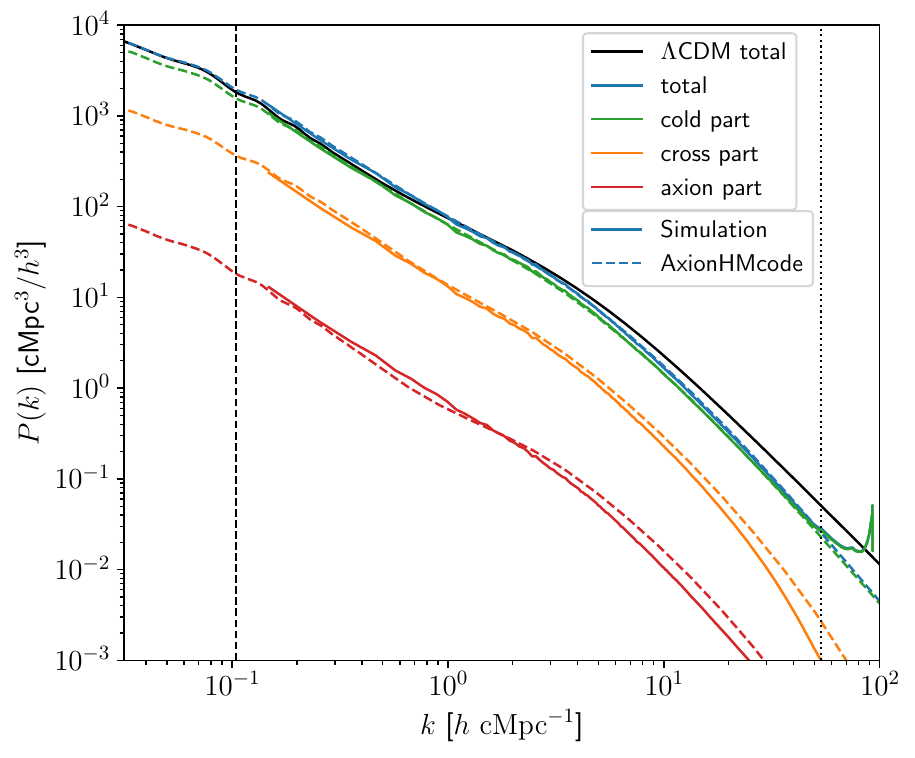}
\caption{\textsc{AxionHMcode} vs MDM simulations. We compare the non-linear power spectrum for our MDM simulation (solid) with the halo model implementation using the new version of \textsc{AxionHMcode} (dashed). Results are shown for $f=0.1$ and $m=3.16\times 10^{-25}$\,eV at redshifts $z=1$. We show the full power spectrum (blue) and the three contributing terms from Eq.~\eqref{e_three_terms} (green, orange, and red). For comparison, we show the total matter power spectrum in a pure $\Lambda$CDM cosmology (black solid). The dashed vertical line indicates the fundamental frequency of the simulation box $k_{\text{f}} = 2\pi/L_{\text{box}}$ while the Nyquist frequency $k_{\text{Ny}} = \pi N_{\text{CDM}}^{1/3}/L_{\text{box}}$ is shown as a vertical dotted line.}
\label{f_psp_illustrate}
\end{figure}

\subsection{Non-Linear Power Spectra}
\label{ss_nonlinear}
We now validate the updated version of \textsc{AxionHMcode}. Recall that the $M_{\text{a}}(M_{\text{c}})$ relation was fitted over the redshift range $z=1-8$, the smoothing parameter was fitted for $z=1-3$, and the remaining \textsc{HMCode-2020} parameters are calibrated up to $z=2$. As a result, \textsc{AxionHMcode} should be used with caution outside the range $z\approx 1-3$ and $f\approx 0-0.3$. In Fig.~\ref{f_psp_illustrate}, we illustrate its predictions for the non-linear power spectrum, compared to results from our MDM simulations for $f=0.1$ at redshift $z=1$. We find very good agreement between the total predicted and simulated matter power spectrum (blue solid and dashed). The moderate suppression of power relative to $\Lambda$CDM on scales $k> 1 \, h \, \text{cMpc}^{-1}$ is well captured. The agreement between \textsc{AxionHMcode} and simulations extends to the cold matter power spectrum (green), though there is a slight discrepancy for the cross and FDM–FDM power spectra (orange and red) at $k> 10 \, h$\,cMpc$^{-1}$. Note that beyond $k> 10 \, h$ cMpc$^{-1}$ we expect strong effects from baryonic physics \citep[see e.g.][]{Mead_2021}, which are not accounted for in either our MDM simulations or \textsc{AxionHMcode}.

\textsc{AxionHMcode} also captures the \textit{enhanced} non-linear power on scales $k \approx (1-10) \, h$\,cMpc$^{-1}$ relative to $\Lambda$CDM. The effect has been reported by \cite{Vogt_2023} who traced it to an enhancement in the cross and FDM–FDM power spectra, which in turn is caused by the coherence of the soliton in the axion halo density profile, increasing the correlation function of the axion field on such scales. The enhancement is stronger for axion masses around $m\approx 10^{-22}$\,eV (shifting toward higher $k$) and was also observed in pure FDM cosmologies when accounting for `quantum pressure' \citep{Nori_2018_2, May_2022}.

\begin{figure*}
\includegraphics[width=\textwidth]{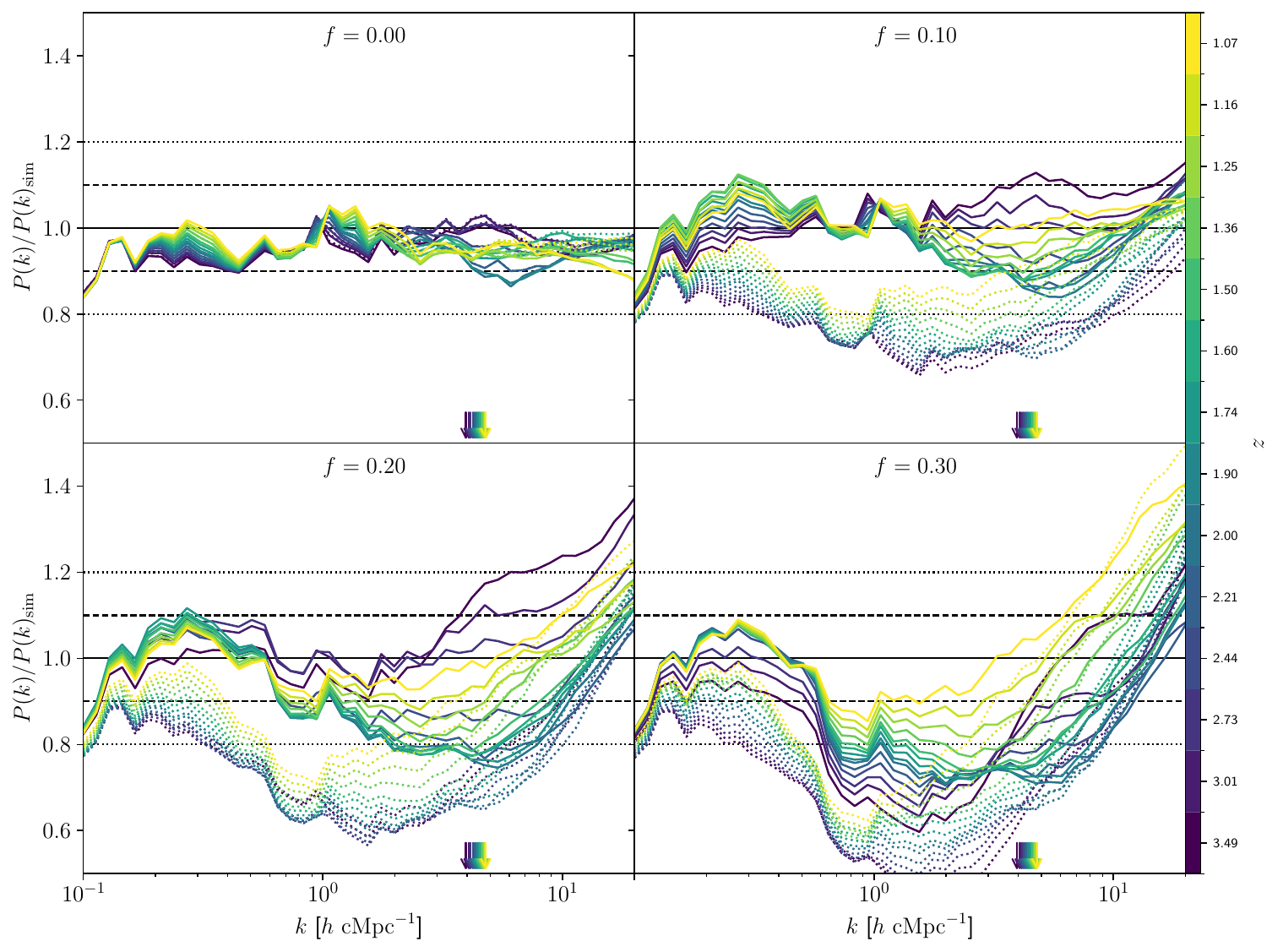}
\caption{Comparison of non-linear matter power spectrum predictions from the updated version of \textsc{AxionHMcode} and MDM simulations at across redshifts $z=1-3.5$. Solid lines represent the new model, while dotted lines show the previous version from \protect\cite{Vogt_2023}. Horizontal dashed and dotted lines indicate 10\,\% and 20\,\% deviations, respectively. Arrows at the bottom mark comoving Jeans scale $k_{\text{J}}$ at each redshift (see Eq.~\ref{e_jeans_fdm}). The axion fraction $f$ increases from the top left panel ($f=0.0$) to the bottom right ($f=0.3$) along with a decrease in overall accuracy of \textsc{AxionHMcode}.}
\label{f_psp}
\end{figure*}

Fig.~\ref{f_psp} displays the non-linear matter power spectrum predictions from the updated version of \textsc{AxionHMcode} at redshifts $z=1-3.5$. We use the default configuration, which includes large-scale damping of the one-halo term, perturbative damping of the two-halo term, halo bloating, and transition smoothing between one- and two-halo terms in the quasi-linear regime. These predictions are compared with MDM power spectra derived from our simulations using \textsc{Pylians} \citep{Villaescusa_2018}.

As anticipated, \textsc{AxionHMcode} demonstrates excellent agreement (deviations less than 10\,\%) with simulations for pure CDM ($f=0.0$) across all scales and redshifts examined. For an axion fraction of $f=0.1$, the code maintains high accuracy on both large (small $k$) and small scales (large $k$) at all considered redshifts, with maximum deviations also remaining below 10\,\%. When the axion fraction is increased to $f=0.2$, maximum deviations rise to approximately 20\,\% for scales with $k< 10 \, h \, \text{cMpc}^{-1}$; however, at redshifts around $z\approx 1$, deviations on these scales remain under 10\,\%. For $f=0.3$, the maximum deviations increase to about 30\,\% for scales with $k< 10 \, h \, \text{cMpc}^{-1}$, though at redshifts around $z\approx 1$, deviations do not exceed approximately 20\,\% on these scales.

We are unable to assess the agreement with MDM simulations for redshifts $z<1$ due to the potential for unresolved high velocity dispersions in the FDM solver (see Sec.~\ref{ss_sim_setup_mdm}). Similarly, neither \textsc{HMCode-2020} nor \textsc{AxionHMcode} are anticipated to perform well at redshifts above $z\approx 3.5$, which defines the upper limit of our redshift window. Note that our MDM simulation data is limited to FDM fractions $f\leq 0.3$. However, this limitation is not critical, as ultralight axions are permitted to exist in substantial portions within the FDM window but not at extremely high fractions close to $f_{\text{max}}$, which are increasingly ruled out with greater significance beyond the FDM window. Additionally, our fitting of \textsc{AxionHMcode} parameters is constrained to $\Lambda$CDM and MDM cosmologies with axion mass $m=3.16\times 10^{-25}$\,eV, which lies within the FDM window $10^{-25}$\,eV $\lesssim m \lesssim 10^{-23}$\,eV. Due to the lack of MDM simulation data for other axion mass values, we cannot validate the accuracy of \mbox{\textsc{AxionHMcode}} for axion masses significantly different from $m=3.16\times 10^{-25}$\,eV, except in the limit as $m \rightarrow \infty$. The exponent value $b=0.0450$ in Eq.~\eqref{e_alpha1} is thus an educated guess based on the expectation of a weak dependence on $m$. Future MDM simulations will provide the opportunity to refine this estimate and validate the parameter $b$.

\section{Conclusions}
\label{s_mdm_outlook}
In view of tighter constraints being put on pure, single-field ultralight axion DM \citep[see][for a compilation]{Dome_2022}, it is promising to relax the requirement that ultralight axions must comprise all of the DM in the Universe. In this work, we focused on the FDM window $10^{-25}$\,eV $\lesssim m \lesssim 10^{-23}$\,eV in which ultralight axions are allowed to exist in large portions (albeit not $f=f_{\text{max}}$) \citep[although see][]{Shevchuk_2023, Lazare_2024, Winch_2024}.

\textit{Methods:} We implemented an MDM gravity solver (see Sec.~\ref{ss_pseudo}) and ran state-of-the-art simulations of mixed ultralight axion cosmologies dominated by CDM ($f<0.5$). Our MDM simulations were designed to capture the wave dynamics across small and intermediate length scales, with particular emphasis on achieving numerical convergence and resolution thresholds. ICs were set up carefully via second order Lagrangian perturbation theory (2LPT) using \textsc{Music} \citep{Hahn_2011} based on matter and velocity power spectra calculated using \textsc{AxionCamb} \citep{Hlozek_2015}. By rigorously enforcing criteria such as velocity resolution and resolving the axion half-mode scale $k_{1/2}$, while also ensuring accurate representation of halo populations, our simulations faithfully reproduce internal halo structures above redshifts $z \approx 1$, providing a robust platform for evaluating common \mbox{(semi-)}analytical techniques such as halo models.

\textit{Halo Mass Distribution:} We identified halos using the \textsc{Rockstar} particle-based halo finder applied on the (equal-mass) CDM distribution and report total halo mass distributions ($M_{\text{tot}} = M_{\text{c}} + M_{\text{a}}$) in MDM. We found good agreement between the Sheth–Tormen model based on the linear MDM matter power spectrum and the inferred HMF across a very wide range of redshifts $z=1-10$ and axion fractions $f=0.0-0.3$, justifying the usage of \mbox{\textsc{Rockstar}} on the CDM component a posteriori. The HMF in MDM cosmologies branches off from the CDM one at the characteristic mass $M_0 = 1.6\times 10^{10}(m/10^{-22} \, \text{eV})^{-4/3} \, M_{\odot}$, but instead of plateauing toward low mass as in WDM or turning over as in pure FDM, the HMF continues to increase in a power-law fashion as $M_{\text{tot}}$ decreases. By providing best-fit results to a one-parameter model of the MDM HMF, Eq.~\eqref{e_nhmf_model}, we hope to facilitate parameter sampling across MDM cosmologies in Bayesian constraint and forecast analyses.

\textit{Density profiles:} We fit the Einasto model against total DM density profiles and found the Einasto parametrisation to be reliable across the entire range of FDM fractions $f=0.0-0.3$ and redshifts $z=1-10$ studied. The resulting median concentration–mass relation $c(M_{\text{tot}})$ is in good agreement with the \cite{Ludlow_2016} model based on extended Press–Schechter theory, suggesting that instead of turning over at around two decades above the half-mode mass, $100\times M_{1/2}$, as in pure WDM and FDM, the concentration exhibits a decrease before recovering and increasing toward smaller halo mass $M_{\text{tot}}$. This results in an effective flat U-shaped $c(M_{\text{tot}})$ relation for high values of the axion fraction $f$, and is in agreement with insights from the halo mass distribution.

\textit{Calibrating AxionHMcode:} We aimed at improving the calibration of the halo model code \textsc{AxionHMcode} based on a biased tracer approach using insights from our MDM simulations. The aforementioned success of reproducing analytical total halo mass distributions and concentration–mass relations based on halos identified solely using the CDM component lends additional credence to the viability of the biased tracer approach. The modifications we introduce (apart from minor ones) are threefold: First, we model the axion halo mass–cold halo mass relation $M_{\text{a}}(M_{\text{c}})$ as a broken power law below the Jeans mass $M_{\text{J}}$ and retain the cosmic mean relation $M_{\text{a}} = (\Omega_{\text{a}}/\Omega_{\text{c}}) M_{\text{c}}$ above $M_{\text{J}}$. Second, we generalise the transition smoothing parameter $\alpha$ to MDM with a dependence on $m$ and $\Omega_{\text{a}}$ while heeding the spill-over effects of strong smoothing on the large-scale regime using a logistic function for the wavenumber-dependent smoothing $\alpha_{\text{c}}(k)$. Third, we introduce various speed-ups by making sure numerical functions are not evaluated too often, leading to a slight increase in memory requirements while reducing run-time on a single-core machine to below $1$ minute for a single evaluation of \textsc{AxionHMcode}. The code exhibits excellent agreement with simulations for pure $\Lambda$CDM, with deviations under 10\,\% on scales below $k< 20 \, h \, \text{cMpc}^{-1}$ and redshifts $z=1-3.5$. For axion fractions $f\leq 0.3$, the model maintains accuracy with deviations under 20\,\% at redshifts $z\approx 1$ and scales $k< 10 \, h \, \text{cMpc}^{-1}$, though deviations can reach up to 30\,\% for higher redshifts when $f=0.3$.

\textit{Outlook:} Mixed ultralight axion cosmologies dominated by CDM ($f<0.5$) have their clustering properties determined to first order by the CDM component. More precisely, the peak height distribution is largely shaped by the cold component as first suggested by \cite{Massara_2014}, motivating not only the very assumptions underlying \textsc{AxionHMcode} but also the identification of halos based solely on the CDM component. MDM models might have the potential to reconcile observational constraints while providing a semi-phenomenological route to understanding the nature of DM. In upcoming forecast and constraint analyses based on observational data, having good control of non-linear predictions in MDM models is key, and this work aims to contribute to that undertaking.

\section{Acknowledgements}
We acknowledge useful discussions with Sophie Vogt. TD acknowledges support from the Isaac Newton Studentship and the UK Research and Innovation (UKRI) Science and Technology Facilities Council (STFC) under Grant No.\ ST/V50659X/1. SM acknowledges support by the National Science Foundation under Grant No.\ 2108931. SB is supported by the UKRI Future Leaders Fellowship (Grant No.\ MR/V023381/1). AL acknowledges support from NASA grant 21-ATP21-0145. DJEM is supported by an Ernest Rutherford Fellowship from the STFC, Grant No. ST/T004037/1. This work used the DiRAC@Durham facility managed by the Institute for Computational Cosmology on behalf of the STFC DiRAC HPC Facility, with equipment funded by BEIS capital funding via STFC capital grants ST/K00042X/1, ST/P002293/1, ST/R002371/1 and ST/S002502/1, Durham University and STFC operations grant ST/R000832/1.

\section{Data Availability}
\label{s_data_availability}
MDM snapshot data and post-processing scripts are made available upon reasonable request.

\bibliographystyle{mnras}
\bibliography{refs}

\label{lastpage}
\end{document}